\documentclass[prd,superscriptaddress,letterpaper]{revtex4-1}

\usepackage{graphicx}
\usepackage{hyperref}
\usepackage{bm}
\usepackage{amsmath}
\usepackage{amssymb}
\usepackage{color}

\newcommand{\hMpc}{\ h^{-1}\text{Mpc}}
\newcommand{\hMpcc}{\ h^{-3}\text{Mpc}^3}
\newcommand{\hGpcc}{\ h^{-3}\text{Gpc}^3}
\newcommand{\ihMpc}{\ h\text{Mpc}^{-1}}

\newcommand{\hMs}{\ h^{-1} M_\odot}
\newcommand{\eh}[1]{\exp{\left[#1\right]}}
\newcommand{\tim}[1]{\times 10^{#1}}

\newcommand{\tr}[1]{\textcolor{red}{#1}}

\newcommand{\derd}{\,\mathrm{d}} 
\newcommand{\ddir}{\delta^\text{(D)}}
\newcommand{\dkron}{\delta^\text{(K)}}
\newcommand{\be}{\begin{equation}}
\newcommand{\ee}{\end{equation}}
\newcommand{\la}{\left\langle}
\newcommand{\ra}{\right\rangle}
\newcommand{\derivd}{\text{d}}
\renewcommand{\vec}{\bm}
\newcommand{\dqc}{\frac{\derivd^3q}{(2\pi)^3}}

\newcommand{\ii}{\text{i}}
\newcommand{\lcdm}{$\Lambda$CDM }

\def\mnras{{MNRAS}}

\def\prd{{PRD}}
\def\jcap{{JCAP}}

\def\apj{{ApJ}}
\def\apjs{{ApJS}}
\def\apjl{{ApJL}}
\def\aap{{A\&A}}
\def\nat{{Nature}}

\def\physrep{{Phys.~ Rep.~}}
\def\aj{{Astronom.~J.}}

\begin{document}
\title{Halo Stochasticity from Exclusion and non-linear Clustering}
\author{Tobias Baldauf}
\email{baldauf@physik.uzh.ch}
\affiliation{Institut f\"ur Theoretische Physik, Universit\"at Z\"urich, Z\"urich, Switzerland}%
\author{Uro\v{s} Seljak}
\affiliation{Institut f\"ur Theoretische Physik, Universit\"at Z\"urich, Z\"urich, Switzerland}%
\affiliation{Physics Department and Lawrence Berkeley National Laboratory, University of California, Berkeley, USA}
\affiliation{Institute for the Early Universe, Ewha University, Seoul, South Korea}
\author{Robert E. Smith}
\affiliation{Max-Planck-Institut f\"ur Astrophysik, Garching, Germany}
\author{Nico Hamaus}
\affiliation{Institut d'Astrophysique de Paris, Universit\'e Pierre et Marie
Curie, Paris, France}
\affiliation{Department of Physics, University of Illinois at Urbana-Champaign, Urbana, IL, USA}
\author{Vincent Desjacques}
\affiliation{D\'{e}partement de Physique Th\'{e}orique \& Center for Astroparticle Physics, Universit\'e de Gen\`eve, Gen\`eve, Switzerland}
\date{\today}
\begin{abstract}
The clustering of galaxies in ongoing and upcoming galaxy surveys contains a wealth of cosmological information, but extracting this information is a non-trivial task since galaxies and their host haloes are stochastic tracers of the non-linear matter density field. 
This stochasticity is usually modeled as the Poisson shot noise, which is constant as a function of wavenumber with amplitude given 
by $1/\bar{n}$, where $\bar{n}$ is the number density of galaxies. 
Here we use dark matter haloes in $N$-body simulations to show evidence for deviations from this simple behaviour and
develop models that explain the behaviour of the stochasticity on large scales.  First, haloes are extended, non-overlapping objects, i.e., their correlation function needs to go to -1 on small scales. This leads to a negative correction to the stochasticity relative to the 
Poisson value
at low wavenumber $k$, decreasing to zero for wavenumbers large compared to the inverse exclusion scale. Second, haloes show a non-linear enhancement of clustering outside the exclusion scale, leading to a positive stochasticity correction. Both of these effects go to zero for high-$k$, making the stochasticity scale dependent even for $k<0.1\ihMpc$. 
We show that the corrections in the low-$k$ regime are the same in Eulerian and Lagrangian space, but that the transition scale 
is pushed to smaller scales for haloes observed at present time (Eulerian space), relative to the initial conditions (Lagrangian space). 
These corrections vary with halo mass and we present approximate scalings with halo mass and 
redshift. We also discuss simple applications of these effects to galaxy samples with non-vanishing satellite fraction, where 
the stochasticity can again deviate strongly from the fiducial Poisson expectation.
Overall these effects affect the clustering of galaxies at a level of a few percent even on very large scales 
and need to be modelled properly if we want 
to extract high precision cosmological information from the upcoming galaxy redshift surveys. 

\end{abstract}
\maketitle

\section{Introduction}
The three dimensional distribution of galaxies has the potential to tell us a lot about the physics governing our Universe.
However, the imprint of the composition and history of our Universe on its structure is usually quantified in terms of the linear power spectrum. From there it is a fair way to go to connect to the distribution of luminous objects. Due to the stochastic nature of the initial conditions the comparison between theory and observation has to be made at a statistical level. Thus, it has become common practice to reduce the data to $n$-spectra and to find a way to push the theory as far as possible in order to make predictions for the observed spectra. This means that the theoretical prediction needs to account  for the fact that galaxies are only sampling the underlying matter distribution. While their distribution is clearly related to the matter distribution, there are a number of distinct features present in the galaxy distribution that are related to their discrete nature and the fact that galaxies form preferentially in high density regions.
\par
Due to the complicated nature of galaxy formation, cosmological constraints from galaxy surveys are usually obtained using a bias model \cite{Kaiser:1984on}. The simplest local bias models \cite{Fry:1992vr} assume a proportionality between the galaxy and matter overdensities. As we will review in detail below in \S \ref{sec:poiss}, the auto power spectrum of a sample of $N$ particles in a volume $V$ is expected to have an additional \emph{scale independent} shot noise component $V/N$. We will refer to this Poisson prediction as fiducial stochasticity. On top of the fiducial Poisson shot noise there are further contributions to the halo power spectrum that are white only over a limited range of wavenumbers and lead to modifications in the $k\to 0$ limit. The latter will be referred to as stochasticity corrections. For instance, the studies of \cite{Seljak:2009ho,Hamaus:2009op} found evidence for a sub-Poissonian noise in the halo distribution in $N$-body simulations (see also \cite{CasasMiranda:2002on,Manera:2011th}) and used this concept to increase the information content extractable from surveys by weighting haloes accordingly. Subsequently, this approach was used to improve constraints on primordial non-Gaussianity \cite{Hamaus:2011op} and redshift space distortions \cite{Hamaus:2012op}.
\par
Thus far, the origin of these stochasticity corrections has not been understood consistently. However, some authors noted that realistic bias models would at some point need to account for the finite size of haloes and the resulting exclusion effects \cite{Sheth:1999bi}.
The effect of halo exclusion on the power spectrum was previously discussed in \cite{Smith:2007sc} in a an Eulerian setting. Here, we will argue that the exclusion effect can not be seen in isolation but has to be combined with the non-linear clustering, which can lead to positive corrections on large scales. This approach partially alleviates the longstanding problem of non-vanishing contributions of the perturbative bias model on the largest scales, where perturbative corrections are considered unphysical. This paper aims at shedding light at the stochasticity properties of halo and galaxy samples and tries to quantify them where possible.
\par
The paper breaks down as follows: We begin in \S \ref{sec:disctrac} with a short review of the standard Poisson shot noise for a sample of discrete tracers. Then, in \S \ref{sec:toymod} we consider some simple toy models to understand the effects of exclusion on the power spectrum, before we go on to discuss more realistic models for the clustering of dark matter haloes in \S \ref{sec:quanti}. In \S \ref{sec:simul} we study the stochasticity and correlation function for a sample of dark matter haloes and a HOD galaxy sample in $N$-body simulations. Finally, we summarize our findings in \S \ref{sec:concl}.

\section{Discrete Tracers}\label{sec:disctrac}

\subsection{Correlation and Power Spectrum}\label{sec:poiss}
The overdensity of discrete tracer particles (dark matter haloes, galaxies etc.) can generically be written as
\begin{equation}
 \delta^\text{(d)}(\vec r)=\frac{n(\vec r)}{\bar n}-1=\frac{1}{\bar n}\sum_i \delta^\text{(D)}(\vec r- \vec r_i)-1,
\end{equation} 
where $\bar n$ is the mean number density of the point-like objects whereas $n(\vec r)$ is their local number 
density. The two-point correlation of this fluctuation field is the expectation value 
\begin{align}
\la\delta^\text{(d)}(\vec r)\delta^\text{(d)}(\vec 0)\ra &=
\frac{1}{{\bar n}^2}\Bigl\langle\sum_{i,j}\ddir\left(\vec r-\vec r_i\right)
\delta^\text{(D)}\left(\vec r_j\right)\Bigr\rangle -\frac{1}{\bar n} 
\Bigl\langle\sum_i\ddir\left(\vec r-\vec r_i\right)\Bigr\rangle-\frac{1}{\bar n} 
\Bigl\langle\sum_j\ddir\left(\vec r_j\right)\Bigr\rangle + 1
\\ 
&= \frac{1}{{\bar n}^2}\delta^\text{(D)}\left(\vec r\right)
\Bigl\langle\sum_i\delta^\text{(D)}\left(\vec r-\vec r_i\right)\Bigr\rangle
+\frac{1}{{\bar n}^2}\Bigl\langle\sum_{i\ne j}\ddir\left(\vec r-\vec r_i\right)
\ddir\left(\vec r_j\right)\Bigr\rangle-1 \nonumber \\
&= \frac{1}{\bar n} \ddir\left(\vec r\right)
+\frac{1}{{\bar n}^2}\Bigl\langle\sum_{i\ne j}\ddir\left(\vec r-\vec r_i\right)
\delta^\text{(D)}\left(\vec r_j\right)\Bigr\rangle-1 \nonumber \\
&= \frac{1}{\bar n}\ddir\left(\vec r\right) + \xi^\text{(d)}(r) \nonumber \;.
\end{align}
We split the sum into an $i=j$ and and $i\neq j$ part, corresponding to the correlation of the discrete particles 
with themselves and the correlation between different particles, respectively.
The second term in the last equality is the reduced two-point correlation function of the tracers. The first term 
arises owing to ``self-pairs'', which are usually ignored in the calculation of real space correlations. Taking 
the Fourier transform of the last expression, the power spectrum of the discrete tracers is
\begin{equation}
P^\text{(d)}(k) = \frac{1}{\bar n} + \int \derd^3 r\, \xi^\text{(d)}(r) \eh{\ii \vec k \cdot \vec r}  \;.
\end{equation}
Self-pairs contribute the usual Poisson white noise $1/\bar n$. The only requirement is that the power spectrum
be positive definite. This implies that the Fourier transform of the two-point correlation $\xi^\text{(d)}(r)$ can 
be anything equal or greater than $-1/\bar n$. In the limit $k\to 0$ in particular, the power spectrum tends
towards
\begin{equation}
P^\text{(d)}(k) \xrightarrow{k\to 0} \frac{1}{\bar n} + \int \derd^3 r\, \xi^\text{(d)}(r) \;,
\end{equation}
where the intregal of $\xi^\text{(d)}(r)$ over the whole space can be positive, zero or negative (but greater than
$-1/\bar n$) depending on the nature of the discrete tracers. This can lead to super-poisson or sub-poisson white 
noise in the low-$k$ limit. 

At $k=0$, the power spectrum is $P^\text{(d)}(0)=0$ because the fluctuation field $\delta^\text{(d)}(\vec r)$ is 
defined relative to the mean number density, hence $\langle\delta^\text{(d)}\rangle =0$. This implies that 
$P^\text{(d)}(k)$ drops precipitously on very large scales (so it must be discontinuous at $k=0$) regardless the value 
of $\int \derd^3 r\, \xi^\text{(d)}(r)$. To convince ourselves that this is indeed the case, we can write the 
Fourier modes of the tracer fluctuation field as
\begin{equation}
\delta^\text{(d)}(\vec k) = \frac{1}{\bar n} \sum_i \eh{\ii \vec k \cdot \vec r_i} - \int \derd^3 r\,
\eh{\ii \vec k \cdot \vec r} \;.
\end{equation}
To calculate $\delta^\text{(d)}(k=0)$ (which is formally the difference between two infinite quantities), we first
assume $N,V\gg 1$ at fixed average number density $\bar n\equiv N/V$, and then take the limit $N,V \to\infty$. 
 We thus have for the Fourier transform of the density field
 \be
\delta^\text{(d)}(\vec k) = \frac{1}{\bar n} \sum_i \eh{\ii \vec k \cdot \vec r_i} - V\dkron_{\vec k,\vec 0} \;,
\label{eq:finitevoldeltak}
 \ee
 which for $\vec k=0$ yields
\begin{equation}
\delta^\text{(d)}(\vec 0)= \frac{V}{N} N - V = 0 \;.
\end{equation}
This obviously holds also for a finite number $N$ of tracers in a finite volume $V$. Therefore, the fact that 
$\int \derd^3 r\, \xi^\text{(d)}(r)$ can be different from zero has nothing to do with the fact that 
$\langle\delta^\text{(d)}\rangle=0$, nor with the so-called ``integral constraint'' that appears when measuring an
excess of pairs relative to a random distribution in a finite volume \cite{Labatie:2010un,Peacock:1991th}.
\subsection{The Effect of Exclusion with Clustering}
Let us now account for the fact that haloes are the centre of an ensemble of particles, which by definition can not overlap, and that these are clustered. Exclusion means it is forbidden to have two haloes closer than the sum of their radii $R$. This fact can be accounted for by writing the correlation function of the discrete tracers as
\be
\xi_\text{hh}^\text{(d)}(r)=
\begin{cases}
-1 & \text{for}\ r<R\\
\xi_\text{hh}^\text{(c)}(r) &\text{for}\ r\geq R,
\end{cases}
\ee
where the fictitious continuous correlation function $\xi_\text{hh}^\text{(c)}(r)$ is defined for $r\in[0,\infty]$ and would for instance be related to the matter correlation function by the local bias model (see \S \ref{sec:biasmodel} below). Enforcing this step at the exclusion radius is certainly overly simplistic, since any triaxiality or variation of radius within the sample will smooth this step out. We will come back to this issue later.
\\
For generic continuous clustering models, we can write the Fourier transform of the correlation function as
\begin{align}
\int_0^\infty \derd^3 r \xi_\text{hh}^\text{(d)}(r) j_0(kr)=&-\int_0^R \derd^3 r  j_0(kr)+\int_R^\infty\derd^3 r \xi_\text{hh}^\text{(c)}(r)j_0(kr)\nonumber\\
=&-V_\text{excl}W_R(k)-\int_0^R \derd^3 r \xi_\text{hh}^\text{(c)}(r)j_0(kr)+\int_0^\infty \derd^3 r \xi_\text{hh}^\text{(c)}(r)j_0(kr)\\
=&-V_\text{excl}W_R(k)-V_\text{excl} \left[W_R*P_\text{hh}^\text{(c)}\right](k)+P_\text{hh}^\text{(c)}(k)\nonumber,
\end{align}
where the exclusion volume is $V_\text{excl}=4\pi R^3/3$, $j_0$ is the zeroth order spherical Bessel function and the Fourier transform of the top-hat window is given by
\be
W_R(k)=3\frac{\sin(kR)-kR \cos(kR)}{(kR)^3}
\ee
and where the notation $[A*B](k)$ describes a convolution integral
\be
[A*B](k)=\int \dqc A(q)B(|\vec k -\vec q|).
\ee
We also defined the continuous power spectrum as the full Fourier transform of the continuous correlation function 
\be
P_\text{hh}^\text{(c)}(k)=\int_0^\infty \derd^3 r \xi_\text{hh}^\text{(c)}(r) j_0(kr).
\ee
Combining the above results with the fiducial stochasticity contribution we finally have for the power spectrum of the discrete tracers
\be
P_\text{hh}^\text{(d)}(k)=\frac{1}{\bar n}+P_\text{hh}^\text{(c)}(k)-V_\text{excl}W_R(k)-V_\text{excl}\left[W_R*P_\text{hh}^\text{(c)}\right](k).
\label{eq:discretepower}
\ee
This equation is the basis of our paper and we will thus explore it in detail.

It is common practice to ignore the exclusion window and to approximate the continuous power spectrum by the linear local bias model, which yields for the power spectrum of the discrete tracers in the Poisson model
\be
P_\text{hh}^\text{(d)}(k)=\frac{1}{\bar n}+b_1^2P_\text{lin}(k).
\ee
This needs to be modified because of exclusion and non-linear effects. 
In practice, for $k>0$, it is difficult to separate the effects. Here we will formally define the 
stochasticity effects discussed in this paper as a stochasticity power spectrum \cite{Hamaus:2009op}
\begin{align}
(2\pi)^3 \ddir(\vec k+\vec k')C(k)=&\Bigl\langle\bigl[\delta_\text{h}(\vec k)-b_1 \delta_\text{m}(\vec k)\bigr]\bigl[\delta_\text{h}(\vec k')-b_1 \delta_\text{m}(\vec k')\bigr]\Bigr\rangle\nonumber\\
=&
(2\pi)^3 \ddir(\vec k+\vec k')\Bigl[P_\text{hh}(k)-2b_1 P_\text{hm}(k)+b_1^2 P_\text{mm}(k)\Bigr],
\label{eq:snmatrixdiag}
\end{align}
where $b_1=P_\text{hm}(k)/P_\text{mm}(k)$ is the first order bias from the cross-correlation in the low-$k$ limit. 
We then have in the low-$k$ limit
\be
P_\text{hh}^\text{(d)}(k)=C(k)+b_1^2P_\text{lin}(k).
\ee
One could make this generally valid at all $k$ by defining $b(k)=P_\text{hm}(k)/P_\text{mm}(k)$, but we will 
not do this here, and instead explore physically motivated models of non-linear bias. 
In this paper we are interested in the stochasticity power spectrum $C(k)$ and in particular its 
limit as $k \rightarrow 0$. 

\begin{figure}[t]
	\centering
	\includegraphics[width=0.49\textwidth]{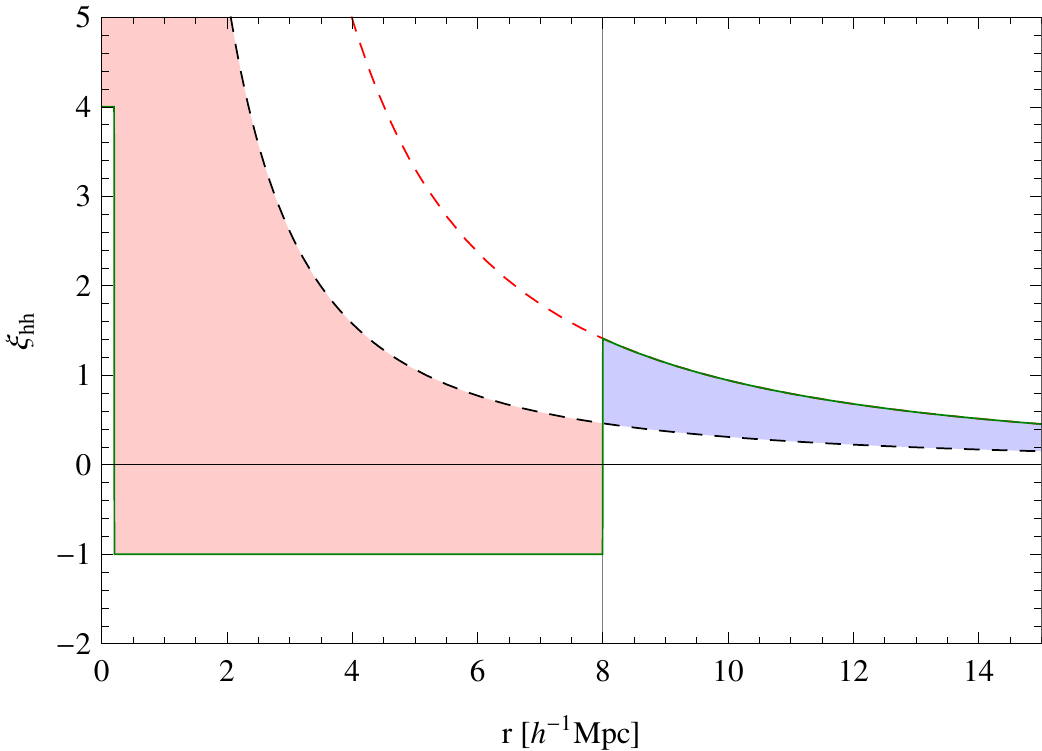}
	\caption{Cartoon version of the correlation function of discrete tracers. Continuous linear correlation function (black dashed) and non-linear correlation function (red dashed). The true correlation function of discrete tracers (green solid line) agrees with the non-linear continuous correlation function outside the exclusion scale and is -1 below, except for the delta function at the origin arising from discreteness. Thus, there are two corrections compared to the continuous linear bias model, a negative correction inside the exclusion radius (red shaded) and a positive one outside the exclusion radius due to non-linear clustering (blue shaded).}
	\label{fig:sketch}
\end{figure}

What are the corrections arising from the exclusion and deviations from the local bias model?
In the low $k$-limit, the window function scales as $W_R(k)\xrightarrow{k\to0}1-k^2 R^2/10$. Hence, the convolution integral leads to a constant term plus corrections scaling as $k^2 R^2$ times moments of the continuous power spectrum
\be
\left[W_R*P_\text{hh}^\text{(c)}\right](k) \xrightarrow{k\to0} \int \dqc P_\text{hh}^\text{(c)}(q)W_R(q)+k^2 R^2 \int \dqc P_\text{hh}^\text{(c)}(q) \left[W_R(q)\left(\frac{1}{(q R)^2}-\frac16\right)-\frac{\sin(q R)}{(qR)^3}\right].
\ee
Thus, irrespective of the shape of the continuous power spectrum on large scales, exclusion always introduces a white ($k^0$) correction on large scales.\\
Fig.~\ref{fig:sketch} illustrates the behaviour of the correlation function of discrete tracers. 
In the very popular local bias model, the clustering of dark matter haloes is modeled at leading order as $P_\text{hh}^\text{(c)}(k)=b_1^2 P_\text{lin}(k)$. In configuration space this leads to $\xi_\text{hh}^\text{(c)}(r)=b_1^2 \xi_\text{lin}(r)$, shown by the black dashed line. 
We will consider this linear bias model as the fiducial model on top of which we define corrections.
Non-linear halo clustering suggests an enhancement proportional to higher powers of the linear correlation function as exemplified by the red dashed line. Our above arguments suggest that this clustering model, if at all, can only be true outside the exclusion radius. Inside this radius the probability to find another halo is zero, leading to $\xi_\text{hh}^\text{(d)}(r<R)=-1$.\\
An intuitive understanding of the corrections can be obtained in the $k\to0$ limit, where the halo power spectrum is given by an integral over the correlation function and can thus be written as
\be
P_\text{hh}^\text{(d)}(k)\xrightarrow{k\to 0}\frac{1}{\bar n}-V_\text{excl}-b_1^2 \int_0^R\derd^3 r\, \xi_\text{lin}(r) +\int_R^\infty \derd^3 r \left[\xi_\text{hh,NL}^\text{(c)}(r)-b_1^2 \xi_\text{lin}(r)\right],
\ee
where we introduced $\xi_\text{hh,NL}^\text{(c)}(r)$ to account for generic non-linear continuous models of the halo clustering.
The red and blue shaded regions in Fig.~\ref{fig:sketch} show the negative and positive corrections with respect to the linear bias model for which we would have in absence of exclusion $P_\text{hh}^\text{(d)}(k)\xrightarrow{k\to 0}1/\bar n$.
Note that the non-linear halo-halo correlation function could in principle be smaller than the linear bias prediction. Our above notion of a positive correction arising from the non-linear correction outside the exclusion radius is solely based on local bias arguments. In general this statement should be relaxed (for an example see App.~\ref{app:peakeffects}) and the blue region could have either sign.

\section{Toy Models}\label{sec:toymod}
To show that the exclusion can indeed lower the stochasticity we perform a simple numerical experiment. We consider a set of hard sphere haloes of radius $R/2$. For this purpose, we distribute $N$ particles randomly in a cubic box ensuring that $\left|\vec x_i-\vec
x_j\right|>R$ for all pairs of particles $(i,j)$.  
The corresponding correlation function is expected to be zero except for scales $r<R$, where $\xi=-1$
due to exclusion. Thus we expect the fiducial stochasticity to be lowered by $4\pi R^3/3$ in the $k \to 0$ limit. 
For an intuitive derivation of the corrections to the power spectrum we will consider a fixed number of particles $N$ in a finite volume $V$.
Using Eq.~\eqref{eq:finitevoldeltak} the auto-power spectrum of the tracer particles can be written as \footnote{{The finite grid leads to $\ddir(\vec k-\vec k')=\frac{V}{(2\pi)^3}\dkron_{\vec k,\vec k'}$ and consequently $V \dkron_{\vec k,\vec k'} P(k)=\la\delta(\vec k)\delta(-\vec k') \ra$.}}
\begin{align}
 P^\text{(d)}(k)=&\frac{1}{V}\Bigl\langle\delta^\text{(d)}(\vec k)\delta^\text{(d)}(-\vec k)\Bigr\rangle\nonumber\\
=&\frac{V}{N^2}\sum_{i=j} \Bigl\langle\eh{\ii \vec k \cdot (\vec r_i-\vec r_j)}\Bigr\rangle+\frac{V}{N^2}\sum_{i\neq j} \Bigl\langle\eh{\ii \vec k \cdot (\vec r_i-\vec r_j)}\Bigr\rangle-V\delta^\text{(K)}_{\vec k,\vec 0}\label{eq:equalsums}\\
=&\frac{1}{\bar n}+\frac{V}{N^2}\sum_{\ii\neq j}\la\eh{\ii \vec k \cdot (\vec r_i-\vec r_j)}\ra-V\delta^\text{(K)}_{\vec k,\vec 0}\nonumber.
\end{align}
This yields for the hard sphere sample, which we consider as a proxy for excluded haloes
\be
P^\text{(d)}_\text{hh}(k)=\frac{1}{\bar{n}_\text{h}}-\frac{4\pi R^3}{3} W_R(k).
\ee
In Fig.~\ref{fig:randomwgal} we show the power spectrum of this toy halo sample for $R=8 \hMpc$, $N=800$ and
$V=300^3 \hMpcc$. 
We clearly see that the measured power follows the exclusion corrected stochasticity.
The window is close to unity on large scales and decays at $k\approx 1/R$, i.e., the fiducial shot noise
is recovered for high $k$. This is a first indication for stochasticity not being scale independent.
Note that the above derivations are only true in the limit, where the total exclusion volume is small compared to the total volume and thus allows for a quasi random distribution (about 0.8\% volume coverage in our case).
\subsection{Satellite Galaxies}
Galaxies are believed to populate dark matter haloes. Let us consider the simple case that each of the dark matter haloes under consideration hosts a central galaxy that, as the name suggests, coincides with the halo centre plus a fixed number $N_\text{s,h}$ of satellite galaxies, such that the total number of satellite galaxies is given by $N_\text{s}=N_\text{s,h}N_\text{h}$.
For simplicity, we will assume that the galaxies are distributed according to a profile $\rho_\text{s}(r)$ with typical scale $R_\text{s}$ around the centers of the host halo centers.
For long wavelength modes $k<1/R_\text{s}$ the $N_\text{s,h}$ galaxies within one halo are effectively one particle, which is why on large scales we expect the stochasticity of the satellite galaxy sample to be equal to the one of the host haloes and only for scales $k>1/R_\text{s}$ the modes can probe the distinct nature of the particles and the stochasticity goes to $1/\bar{n}_\text{s}$. 
\par
We can evaluate our model Eq.~\eqref{eq:equalsums} to obtain the satellite-satellite power spectrum
\begin{align}
P^\text{(d)}_\text{ss}(k)=&\frac{V}{N_\text{s}}+\frac{V}{N_\text{s}^2}\sum_{\text{h}_i} \sum_{\text{s}_j\in \text{h}_i}\sum_{\text{s}_l\neq \text{s}_j\in \text{h}_i}\la\eh{\ii \vec k\cdot (\vec r_j -\vec r_l)}\ra+\frac{V}{N_\text{s}^2}\sum_{\text{h}_i} \sum_{\text{s}_j\in \text{h}_i} \sum_{\text{h}_m\neq\text{h}_i} \sum_{\text{s}_l \in \text{h}_m} \la\eh{\ii \vec k\cdot (\vec r_j-\vec r_l)}\ra\nonumber\\
=&\frac{V}{N_\text{s}}+\frac{V}{N_\text{s}}(N_\text{s,h}-1)\la\eh{\ii k R_\text{s}\mu}\ra^2-\frac{4\pi R^3}{3}u^2_\text{s}(k) W_R(k)\nonumber\\
=&\frac{1}{\bar{n}_\text{s}}\bigl[1+(N_\text{s,h}-1)u_\text{s}^2(k)\bigr]-\frac{4\pi R^3}{3}u_\text{s}^2(k) W_R(k)
\end{align}
Here $\mu$ is the cosine of the angle between $\vec k$ and $\Delta \vec r_{ij}=\vec r_i -\vec r_j$ that is averaged over, $u_\text{s}(k)$ is the normalized Fourier transform of the galaxy profile $\rho_\text{s}(r)$. For definiteness we will assume a delta function profile $\rho(r)=\ddir(r-R_\text{s})/r^2$ corresponding to $u_\text{s}(k)=j_0(k R_\text{s})$, where $j_0$ is the zeroth order spherical Bessel function. The two terms in the above equation correspond to the one and two halo terms in the halo model \cite{Seljak:2000an,Cooray:2002ha}, the profile and the fiducial shot noise arise from correlations between particles in the same halo, whereas the exclusion term is dominated by the correlation between distinct haloes.
The results of the numerical experiment are shown in Fig.~\ref{fig:randomwgal} as the green points. The model prediction is shown as the green solid line and describes the simulation measurement very well. On small scales the power is dominated by the fiducial galaxy shot noise and on large scales the host halo stochasticity dominates
\begin{align}
P_\text{ss}^\text{(d)}(k\ll 1/R_\text{s},1/R)=\frac{1}{\bar n_\text{h}}-\frac{4\pi R^3}{3},
&&
P_\text{ss}^\text{(d)}(k\gg 1/R_\text{s},1/R)=\frac{1}{\bar n_\text{s}}.
\end{align}
While the distribution of satellite galaxies on a sphere of fixed radius around the halo centre is very peculiar and unrealistic, the qualitative behaviour is the same for all profiles with finite support.
In the case studied above, the corrections to the fiducial galaxy shot noise $1/\bar{n}_\text{h}$ are always positive. This is due to the high satellite fraction. 
As we will discuss below, this behaviour might be completely different for galaxy samples with small satellite fraction, where the exclusion effect can be more important than the enhancement due to the satellites.
\subsection{Central and Satellite Galaxies}
\begin{figure}[t]
\includegraphics[width=0.49\textwidth]{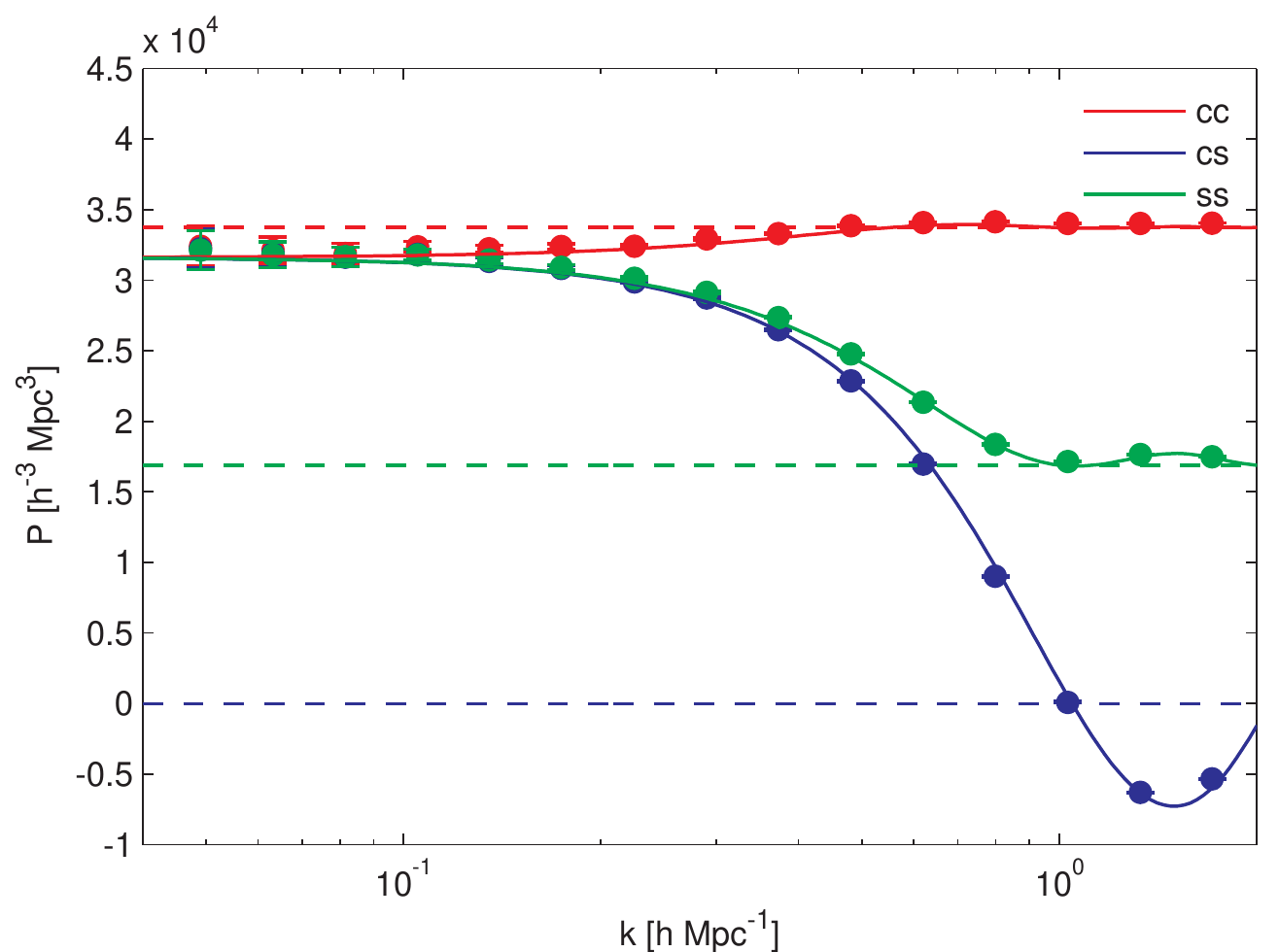}
\caption{Power spectrum of a randomly distributed halo sample obeying exclusion (red points) and corresponding model with (red solid line) and without (red dashed line) exclusion. In a second step we populate these haloes with $N_\text{gal}=2$ satellite galaxies, and calculate the auto power spectrum of the satellite galaxies (green points) and their cross power spectrum with the halo centers (blue points). The blue and green solid lines show our model predictions, whereas the dashed lines show the naive expectation of Poisson shot noise.
}
\label{fig:randomwgal}
\end{figure}

We can also consider the cross power spectrum between halo centers (central galaxies) and the satellite galaxies. In this case there is no Poisson shot noise, since the samples are non-overlapping and the power is dominated by a one-halo term describing the radial distribution of the satellites around the halo centre
\begin{align}
P_\text{cs}^\text{(d)}(k)=&\frac{V}{N_\text{h}N_\text{s}}\sum_{\text{h}_i}\sum_{\text{s}_j\in \text{h}_i}\la \eh{\ii \vec k \cdot (\vec r_j-\vec r_i)}\ra+\frac{V}{N_\text{h}N_\text{s}}\sum_{\text{h}_i}\sum_{\text{h}_j\neq \text{h}_i}\sum_{\text{s}_l\in \text{h}_j}\la\eh{\ii \vec k\cdot (\vec r_{l}-\vec r_i)}\ra\nonumber\\
=&\frac{1}{\bar{n}_\text{h}}u_\text{s}(k)-\frac{4\pi R^3}{3} u_\text{s}(k)W_R(k).
\end{align}
Here we have again a one halo term arising from correlations of the halo center with satellites in the same halo and a two halo term arising from the correlation of satellite galaxies in one halo with the center of another halo. The comparison with the result of the numerical experiment in Fig.~\ref{fig:randomwgal} shows very good agreement.\\
The above discussion is overly simplified as we assume all haloes to be of the same mass and to host the same number of galaxies. Any realistic galaxy sample will be hosted by a range of halo masses (i.e. a range of exclusion radii) and the number of galaxies per halo will also be a function of mass.\\
The total galaxy power spectrum of the combined central and satellite samples can be obtained as a combination of the central-central, central-satellite and satellite-satellite contributions
\be
P_\text{gg}(k)=(1-f_\text{s})^2P_\text{cc}(k)+2f_\text{s}(1-f_\text{s})P_\text{cs}(k)+f_\text{s}^2 P_\text{ss}(k),
\label{eq:weightedsumgg}
\ee
where $f_\text{s}=N_\text{s}/(N_\text{c}+N_\text{s})$ is the satellite fraction.
For realistic satellite fractions for SDSS LRGs \cite{Eisenstein:2001sp} $f_\text{s}\approx0.1$, the weighting of the central-central power spectrum dominates over the contributions from the central-satellite and satellite-satellite power spectra by factors of $9$ and $81$, respectively. A more realistic galaxy sample based on a HOD population of dark matter haloes in a $N$-body simulation will be discussed in Sec.~\ref{sec:realgal}.
\subsection{Toy Model with Clustering}
\label{ssec:toyclust}
Haloes are clustered, i.e., there is an enhanced probability to find two collapsed objects in the vicinity of each other to finding them widely separated. Let us discuss the influence of this phenomenological result on our toy model. For the sake of simplicity let us assume that haloes always come in pairs, i.e., that there is a second halo outside the exclusion scale at typical separation $R_\text{clust}$. This will similarly to satellite galaxies residing in one halo, lead to a \emph{positive} $k^0$ term on large scales, that decays for $k>1/R_\text{clust}$.  In a more realistic setting, not all haloes will come in pairs, some of them will be single objects, others will come in clusters of $n$-haloes. Furthermore not all of them will be separated by exactly the clustering scale.
\par
Some authors argued that any large scale $k^0$-behavior in the perturbation theory description of biasing is unphysical and should be suppressed by constant but aggressive smoothing \cite{Roth:2011te} or by a $k$-dependent smoothing \cite{Chan:2012ha}.  Based on the above considerations, we argue that such terms are just a result of the clustering of haloes and thus \emph{not} unphysical. Whether the magnitude of these effects can be covered by a perturbative treatment such as second order bias combined with perturbation theory, is a different question, which we will pick up later in \S \ref{sec:clustexcl}.
For now, let us note that such a $k^0$ term is also predicted by biasing models that go beyond the local bias model as for instance the correlation of thresholded regions as discussed briefly in the next subsection and for instance in \cite{Beltran:2011ef}.
Generally the clustering scale exceeds the exclusion scale and thus one should expect that the enhancement due to clustering decays at lower $k$ than the suppression due to exclusion. This is actually what happens in the simulations as we will show in \S \ref{sec:simul}.
\subsection{Density Threshold Bias}
\begin{figure}[t]
\includegraphics[width=0.48\textwidth]{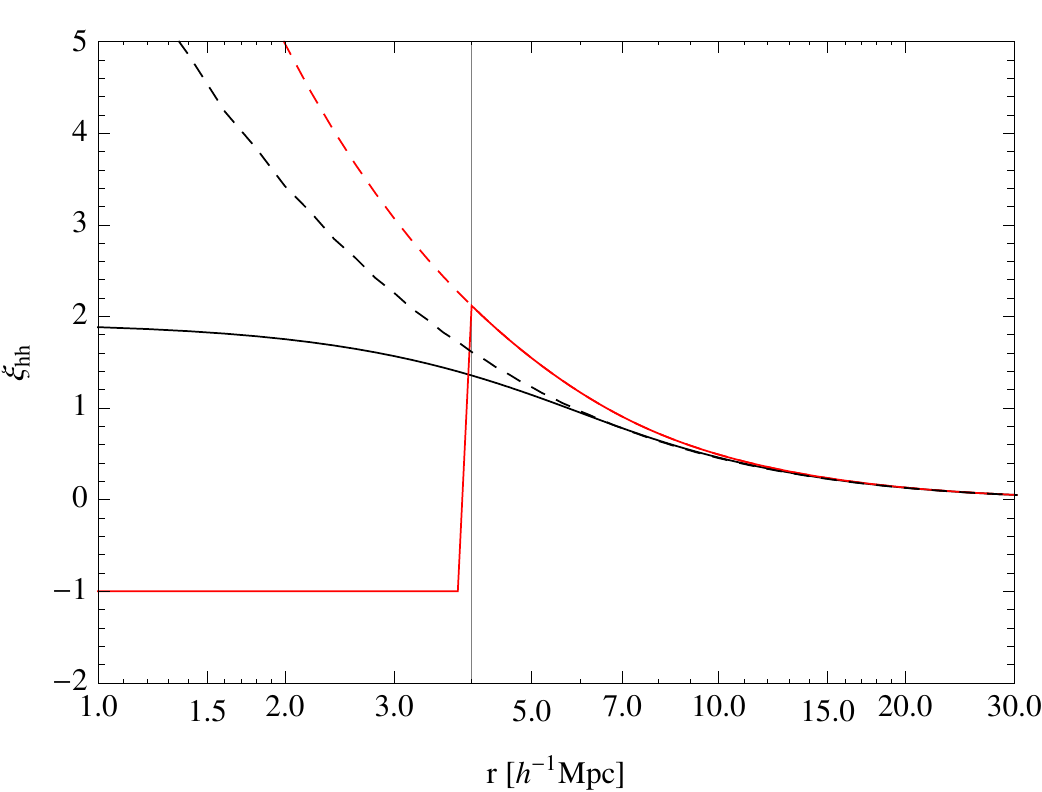}
\includegraphics[width=0.50\textwidth]{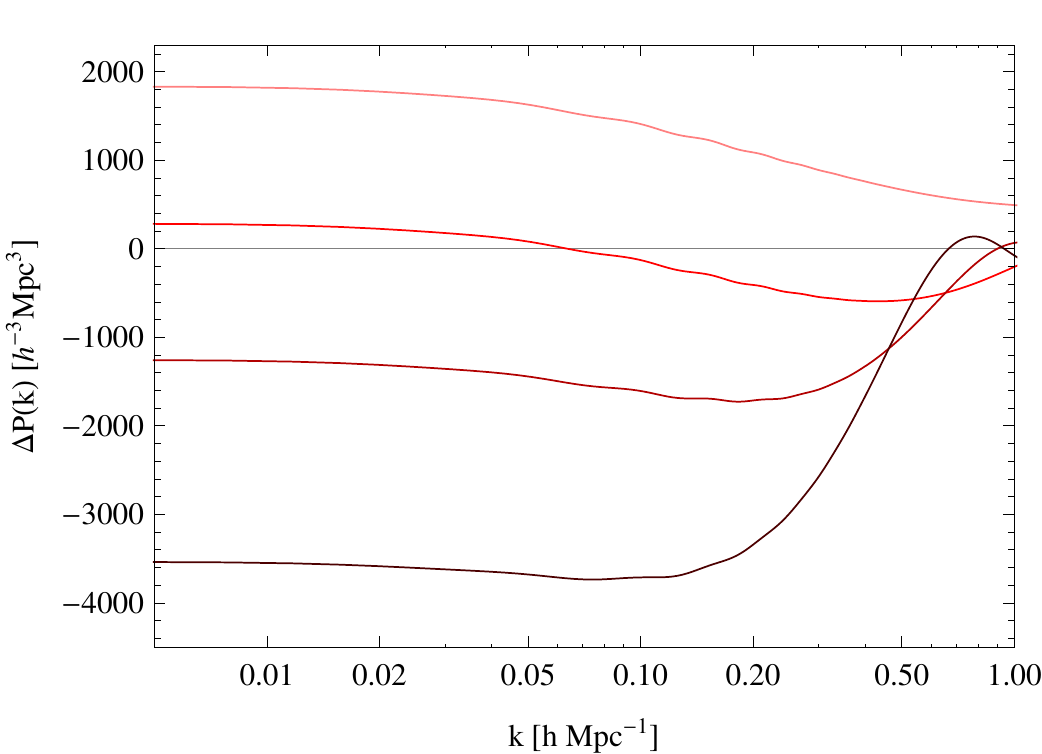}
\caption{Kaiser bias \cite{Kaiser:1984on} in configuration and Fourier space. \emph{Left panel: }Un-smoothed (black dashed) and $R=4\hMpc$ smoothed (black solid) linearly biased matter correlation functions $b_{1,\text{tr}}^2\xi(r)$ and continuous correlation function of the thresholded regions $\xi_\text{tr}(r)$ (red dashed). The red solid line shows a simple implementation of exclusion imposed on the correlation function of the thresholded regions. \emph{Right panel: }Power spectrum correction arising from the non-linear biasing (top line) and effect of increasing exclusion for $R=0,4,6,8 \hMpc$ from top to bottom.}
\label{fig:thresholded}
\end{figure}
The spherical collapse model suggests that spherical Lagrangian regions exceeding the critical collapse density $\delta_\text{c}\approx 1.686$ segregate from the background expansion and form gravitationally bound objects. Hence, the clustering statistics of regions above threshold can tell us something about the clustering of dark matter haloes and galaxies. The study of \cite{Kaiser:1984on} considered the correlation function of regions whose density exceeds a certain value in a Gaussian random field, smoothed with a top-hat window of scale $R$. At the same time this paper pioneered bias models, which are nothing but a large scale expansion of the full correlation function of thresholded regions. Let us see how non-linear or non-perturbative clustering can affect the power spectrum on the largest scales in the full model.\\
The root mean square overdensity within the smoothed regions is given by 
\be
\sigma_R^2=\frac{1}{2\pi^2}\int \derd k\, k^2 W_R^2(k)P_\text{lin}(k).
\ee
The correlation of thresholded regions can be calculated exactly employing the two point probability density function of Gaussian random fields. For simplicity, we will consider regions of a fixed overdensity rather than regions above threshold. The peak height $\nu=\delta_\text{c}/\sigma_R$ can be chosen based on the spherical collapse argument. 
For the correlation function of regions of fixed overdensity one obtains \cite{Beltran:2011ef,Kaiser:1984on}
\be
1+\xi_\text{tr}(r)=\frac{1}{\bar n_\text{tr}^2}
\frac{1}{(2\pi)\sqrt{1-\xi_R^2(r)/\sigma_R^4}}\exp{\left[-\nu^2 \frac{1-\xi_R(r)/\sigma_R^2}{1-\xi_R^2(r)/\sigma_R^4}\right]},
\label{eq:kaiserbias}
\ee
where $\bar n_\text{tr}=1/\sqrt{(2\pi)\sigma_R^2}\eh{-\nu^2/2}$. Here $\xi_R(r)$ is the linear correlation function smoothed on scale $R$.
In the large distance, small correlation limit, the correlation function of the thresholded regions can be approximated by a linearly biased version of the linear correlation function $\xi_\text{tr}(r)=b_{1,\text{tr}}^2 \xi_R(r)$ with $b_{1,\text{tr}}\approx\nu/\sigma_R$. If one is interested in an accurate description of the non-perturbative correlation function on smaller scales, higher orders in the expansion need to be considered. Comparing the expansion of the correlation of thresholded regions in powers of the smoothed correlation to the full non-perturbative result in Eq.~\eqref{eq:kaiserbias}, we can investigate the convergence properties of the linear bias model. The left panel of Fig.~\ref{fig:thresholded} shows the correlation function of thresholded regions for a Gaussian random field smoothed on $R=4\hMpc$ and the linearly biased versions of the smoothed and un-smoothed linear correlation functions.\\
We can now Fourier transform the correlation function of thresholded regions and subtract out the linearly biased power spectrum to obtain the correction introduced by the non-linear clustering
\be
\Delta P_\text{tr}(k)=\text{FT}[\xi_\text{tr}](k)-b_{1,\text{tr}}^2 P_\text{lin}(k).
\ee
As we show in Fig.~\ref{fig:thresholded}, there is a non-vanishing correction in the $k\to0$ limit that is approximately constant on large scales and goes to zero on small scales. The presence of such a correction was discussed in a slightly different context in \cite{Beltran:2011ef}. \\
In its original form, the thresholded regions are a continuous field and thus do not include any exclusion. One could however imagine that the patches defining the smoothing scale do not overlap. In this case the correlation function of thresholded regions should go to $-1$ for $r<2R$.
To show how the exclusion scale affects the correction in the power spectrum we consider a few exclusion radii smaller than two smoothing radii.
As obvious in Fig.~\ref{fig:thresholded}, increasing the smoothing scale first reduces the scale independent correction on large scales, compensates it completely and eventually leads to a negative scale independent correction for $r=2R$. 

\begin{figure}[t]
\includegraphics[width=0.49\textwidth]{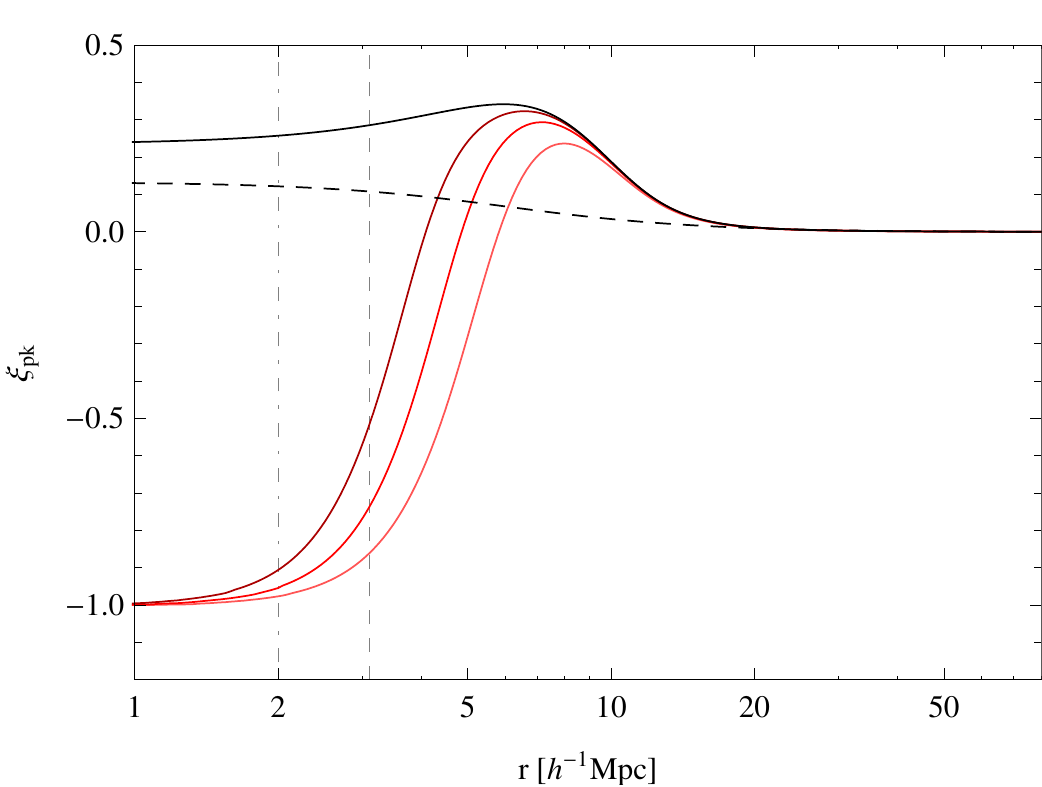}
\includegraphics[width=0.49\textwidth]{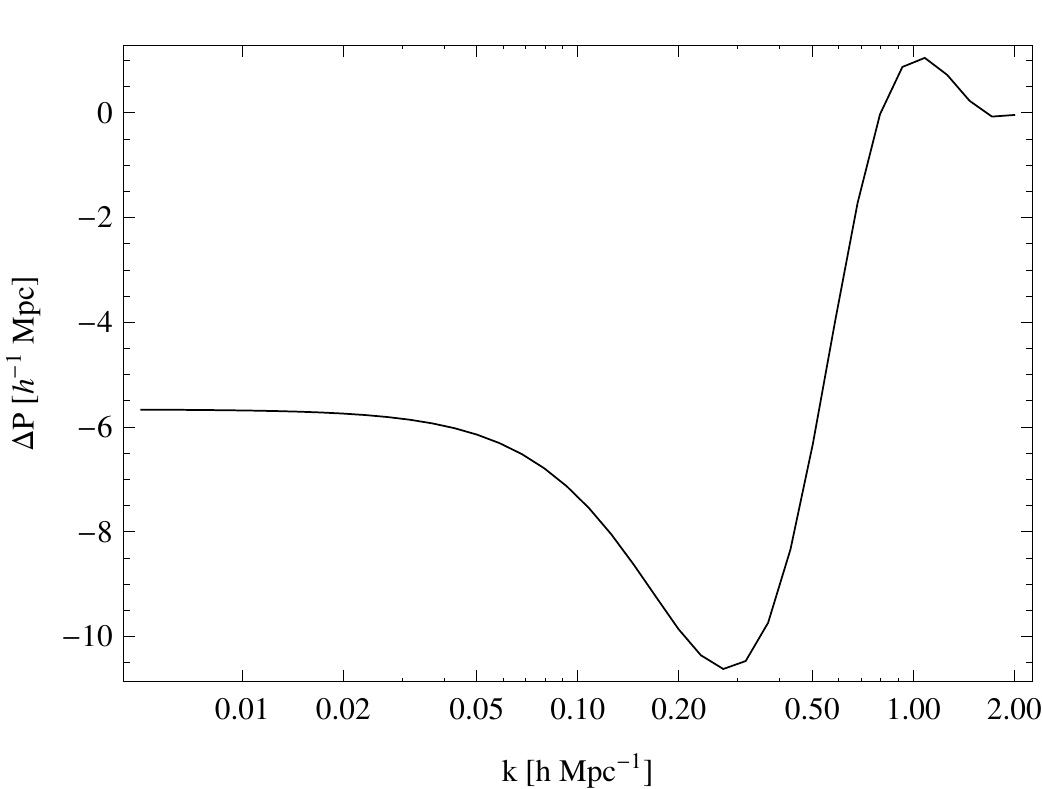}
\caption{Clustering of peaks in a one dimensional skewer through a density field smoothed with a Gaussian filter of scale $R=2\hMpc$ ($M\approx 8.6\tim{12}\hMs$). \emph{Left panel: }For fixed peak height the correlation function flattens out on small scales (black), but with increasing bin width the exclusion becomes stronger. The width of the bin in peak height increases from dark to light red. The linear local bias expansion is the same for all of these models and is shown by the dashed line. For reference we overplot the Gaussian smoothing (dash-dotted) and the top-hat smoothing scale containing the same mass (dashed). \emph{Right panel: }Corresponding stochasticity correction  $\Delta P_\text{pk}(k)=\text{FT}[\xi_\text{pk}](k)-b_{1,\text{pk}}^2 P_\text{lin}(k)$ for the fiducial bin width.}
\label{fig:peaks}
\end{figure}
\subsection{Peak Bias Model}\label{sec:peaks}
While the thresholded regions provide a continuous bias model, the peak model \cite{Bardeen:1985tr,Peacock:1985th} goes beyond in identifying a discrete set of points and providing the correlation function of these points. Most studies of the peak model to date have focused on the large separation limit \cite{Desjacques:2010mo,Desjacques:2008ba,Desjacques:2010re}, where closed form expressions for the peak correlation in terms of the underlying linear correlation function and its derivatives are possible. 
However, \cite{Lumsden:1989th} calculated the one dimensional peak correlation function for a set of power law power spectra and
\cite{Heavens:1999th} computed the two dimensional peak correlation function for peaks in the CMB. The reason for the restriction to one or two dimensions owes to the dimensionality of the covariance matrix that needs to be inverted for the calculation of the peak correlation. In a one dimensional field the covariance matrix is a six by six matrix (field amplitude, first and second derivative at two points).\\
Here we consider realistic \lcdm power spectra in three dimensions, smooth them on a realistic Lagrangian scale $R=2\hMpc$ and evaluate the exact non-perturbative one dimensional correlation of peaks following the approach of \cite{Lumsden:1989th} (see App.~\ref{app:1dpeaks} for a brief review). Note that the correlation of field derivatives diverges for top-hat smoothing, which is why we follow common praxis and employ a Gaussian smoothing. The Gaussian smoothing makes the correlators of field derivatives well behaved but beyond that there is no physical motivation to employ this filter. We study a range of peak heights $\nu$ and also a range of bin widths in $\nu$. The peak correlation function for four different bin widths is shown in the left panel of
Fig.~\ref{fig:peaks}. \\
The first remarkable observation is that peaks of a fixed height don't seem to obey exclusion, only after considering a finite width in peak height, we can observe that the correlation function goes to $-1$ on small scales. The transition scale to the fixed peak height case increases with bin with, i.e., wider bins have a larger exclusion region. 
As above for the thresholded regions we can expand the peak correlation function in the large distance limit and obtain a bias expansion that has contributions from the underlying matter correlation function \emph{and} correlation functions of the derivatives. Doing that, it becomes obvious, that the linear matter bias is only assumed outside of the BAO scale and that there is a distinct scale dependent bias that is partially described by the derivative terms in the bias expansion. Fourier transforming the full peak correlation function and subtracting out the linear biased power spectrum we obtain the correction shown in the right panel of Fig.~\ref{fig:peaks}. 
The qualitative behaviour agrees with the result obtained above for the thresholded sample with an ad hoc exclusion scale. On large scales there is a combination of clustering and exclusion effects, the clustering decays first and then also the exclusion correction goes to zero.
Note that the projected one dimensional matter power spectrum scales as $k^0$ one large scales and thus doesn't vanish in the $k\to0$ limit. This fact makes the distinction between clustering and stochasticity terms in the one dimensional peak model very difficult. We hope to report on results for the full three dimensional peak model in the near future.
\section{Quantifying the Corrections}\label{sec:quanti}
Let us now try to quantify the stochasticity corrections for a realistic halo sample. We expect the effect to be time independent in the 
$k \rightarrow 0$ limit if the same sample of particles is evolved under gravity. Thus, to minimize the influence of non-linearities, we will consider the protohaloes in Lagrangian space. In numerical studies of the effect, we will later define the protohalo as the initial ensemble of particles that form the Friends-of-Friends (FoF) haloes in our final output at redshift $z_\text{f}=0$.

\subsection{Continuous Halo Power Spectrum from Local Bias}\label{sec:biasmodel}
The local Lagrangian bias model assumes that the initial halo density field can be written as a Taylor series in the matter fluctuations at the same Lagrangian position $\vec q$
\be
\delta_\text{h}(\vec q,\eta_\text{i})=b_1^\text{(L)}(\eta_\text{i}) \delta(\vec q,\eta_\text{i}) +\frac{b_2^\text{(L)}(\eta_\text{i})}{2!} \delta^2(\vec q,\eta_\text{i}) +\frac{b_3^\text{(L)} (\eta_\text{i})}{3!}\delta^3(\vec q,\eta_\text{i})+\ldots\ ,
\ee
here $\eta_i$ is the conformal time of the initial conditions and $\vec q$ is the Lagrangian coordinate.
We will follow the approach of \cite{McDonald:2006cl} where the smoothing scale is an unobservable scale, which should not affect $n$-point clustering statistics on scales exceeding the smoothing scale. The above model can be used as the starting point for a coevolution of haloes and dark matter, which finally leads to a Eulerian bias prescription. Recently, such a calculation was shown to correctly predict non-local Eulerian bias terms \cite{Baldauf:2012ev,Chan:2012gr}.
The peak-background-split (PBS) \cite{Mo:1996an} makes predictions for the Lagrangian bias parameters in the above equation, and the corresponding late time Eulerian local bias parameters can then be obtained based on the spherical collapse model. There is some evidence that the peak model yields a better description of some aspects of the initial halo clustering than the local Lagrangian bias model. While we will briefly discuss these effects in App.~\ref{app:peakeffects}, we refrain from using this model for the modelling of the stochasticity corrections, since the the peak bias expansion beyond leading order has not been studied in great detail and its implementation goes beyond the scope of this study.
\par
For the continuous power spectrum in the initial conditions the local Lagrangian bias model predicts
\be
P_\text{hh}^\text{(c)}(k,\eta_\text{i})=\left(b_1^\text{(L)}\right)^2D^2(\eta_\text{i}) P_\text{lin,0}(k)+\frac{1}{2} \left(b_2^\text{(L)}\right)^2 D^4(\eta_\text{i}) I_{22}(k),
\ee
where $D(\eta)$ is the linear growth factor and the scale dependent bias correction is described by
\begin{align}
I_{22}(k)=\int \dqc P_\text{lin,0}(q)P_\text{lin,0}(\left|\vec k-\vec q\right|).
\end{align}
This term leads to a positive $k^0$ contribution in the low-$k$ regime.
In this sense it deviates from typical perturbative contributions to the power spectrum, which start to dominate on small scales.
For this reason, this term was partially absorbed into the shot noise by \cite{McDonald:2006cl}. We will explicitly consider the term, since it describes the effect of non-linear clustering and is responsible for super-Poissonian stochasticity.
The cross power spectrum between haloes and matter is given by
\be
P_\text{hm}^\text{(c)}(k,\eta_\text{i})=b_1^\text{(L)}(\eta_\text{i}) D^2(\eta_\text{i}) P_\text{lin,0}(k)
\ee
and does obviously not contain any second order bias corrections. This statement remains true if higher order biasing schemes are considered, since higher order biases only renormalise the bare bias parameters \cite{McDonald:2006cl}.
\par
Truncating the bias expansion is only valid if $\la\delta^2\ra\ll 1$, which is certainly satisfied on large scales in the initial conditions, but not necessarily on the scales relevant for halo clustering outside the exclusion radius. 
On these scales one might have to consider all the higher order local bias parameters. It is beneficial to calculate this effect in configuration space where the local bias model leads to a power series in the linear correlation function
\be
\xi^\text{(c)}(r)=\sum \frac{\left(b_i^\text{(L)}\right)^2}{i!} D^{2i}(\eta_\text{i}) \xi^i_\text{lin}(r)\xrightarrow{\nu \to \infty} \eh{\frac{\nu}{\delta_\text{c}}\xi_\text{lin}(r)}.
\ee
The limit applies only in the high peak limit and Press-Schechter bias parameters \cite{Press:1974fo}. 
We will restrict ourselves to the quadratic bias model since it can account for the main effects and since using higher order biasing schemes also requires more parameters to be determined. The Press-Schechter and Sheth-Tormen prescriptions provide a rough guideline for the scaling of bias with mass and redshift, but fail to provide correct predictions for the bias amplitude. Thus we obtain the bias parameters from fits to observables not affected by stochasticity (such as the halo-matter cross power spectrum) and obtaining higher order biases would require higher order spectra such as the bispectrum.
Since our discussion is mostly in Lagrangian space we will drop the superscripts E and L from now on and absorb the growth factors into the linear power spectra and correlation functions.

\subsection{Theory Including Clustering and Exclusion}\label{sec:clustexcl}
We can now use the bias model introduced above to evaluate the discrete power spectrum Eq.~\eqref{eq:discretepower}.
We have
\be
P^\text{(d)}(k)=\frac{1}{\bar{n}}+b_1^2P_\text{lin}(k) +\frac12 b_2^2I_{22}(k)-b_1^2 V_\text{excl}[W_R*P_\text{lin}](k)-\frac12 b_2^2 V_\text{excl}[W_R*I_{22}](k)-V_\text{excl} W_R(k).
\ee
The splitting of $I_{22}$, the non-linear clustering term arising from $b_2$, is somewhat counterintuitive, since we expect this term to be active only outside the exclusion scale. Furthermore, there is no corresponding term in the halo-matter or matter-matter power spectra that would cancel the continuous $I_{22}$. Thus we combine the continuous part and the exclusion correction for the non-linear clustering term into a positive correction whose small scale contributions have been removed.
\be
P^\text{(d)}(k)=\frac{1}{\bar{n}}+b_1^2P_\text{lin}(k) +\frac12 b_2^2 I_{22}(k,R)-b_1^2 V_\text{excl}[W_R*P_\text{lin}](k)-V_\text{excl} W_R(k).
\label{eq:sncorrlocbias}
\ee
Here we defined the correction term
\be
I_{22}(k,R)=\int_R^\infty \derd^3 r\  \xi^2(r) j_0(kr).
\ee
A simpler version of Eq.~\eqref{eq:sncorrlocbias} has been presented in \cite{Smith:2007sc}, where the non-linear clustering is neglected and the results are presented in Eulerian rather than Lagrangian space.

In the $k\to 0$ limit the Fourier transform simplifies to a spatial average over the correlation function
\be
P^\text{(d)}(k)\xrightarrow{k\to0}\frac{1}{\bar{n}}+\frac12 b_{2}^2 \int_R^\infty \derd^3 r\ \xi^2(r)-b_{1}^2\int_0^R \derd^3 r\ \xi(r)
-V_\text{excl},
\ee
where the linear bias term vanishes due to $P_\text{lin}(k)\xrightarrow{k\to0}0$.
The fact that the integral over $\xi^2$ runs only from the exclusion scale to infinity mitigates the smoothing dependence of the correction, since smoothing on the scale of the halo affects the correlation function only on the halo scale, which is by definition smaller than the exclusion scale.
\renewcommand{\tabcolsep}{0.2cm}
\begin{table}[t]
\begin{tabular}{lccccc}
\hline
\hline
	bin&$M$&$R_\text{excl}^\text{(L)}$ & $b_1^\text{(L)}$	&$b_2$& $b_1^\text{(E)}$ \\
\hline
I&$1.14\times 10^{13}$&$2.6$&$7.80$&$1174.5$&$1.19$\\ 
II&$1.27\times 10^{13}$&$2.8$&$8.49$&$1206.0$&$1.22$\\ 
III&$1.42\times 10^{13}$&$2.8$&$9.24$&$1287.3$&$1.25$\\ 
IV&$1.62\times 10^{13}$&$3.0$&$10.70$&$1342.9$&$1.27$\\ 
V&$1.89\times 10^{13}$&$3.6$&$13.17$&$1357.7$&$1.31$\\ 
VI&$2.25\times 10^{13}$&$3.9$&$14.94$&$1404.2$&$1.38$\\ 
VII&$2.80\times 10^{13}$&$4.2$&$16.65$&$1635.7$&$1.45$\\ 
VIII&$3.72\times 10^{13}$&$4.9$&$21.92$&$1587.5$&$1.58$\\ 
IX&$5.65\times 10^{13}$&$5.8$&$29.42$&$1439.8$&$1.78$\\ 
X&$1.66\times 10^{14}$&$8.7$&$56.94$&$1346.5 \ii$&$2.54$\\ 

\hline
\hline
\end{tabular}
\caption{Mean masses, exclusion radii, first and second order Lagrangian and Eulerian bias parameters for our $z=0$ halo sample.
Masses are in units of $\hMs$  and radii in units of $\hMpc$. The mass dependence of these parameters is also plotted in 
Fig.~\ref{fig:biasparameters}. Note that the second order bias parameter is just a phenomenological fitting parameter used to get a reasonable representation of the correlation function. We do not claim that our fitting procedure yields an accurate second order bias parameter for the sample. For this purpose on has to employ the bispectrum, which yields second order bias parameters that are in much better agreement with the peak-background split expectation but fail to describe the correlation function.}
\end{table}

\subsection{Stochasticity Matrix}
We will now consider the power spectrum for a set of non-overlapping halo mass bins. We will consider their auto-power spectra and cross power spectra between different halo mass bins $i$ and $j$ and denote this quantity $P_{ij}$, whereas the cross power between a certain mass bin and the matter is denoted $P_{i\delta}$.
The sum over equal pairs in Equation \eqref{eq:equalsums} is only present for the auto power spectra and thus the  $1/\bar n$ shot noise affects only the diagonal entries of the power spectrum matrix $P_{ij}(k)$. On the other hand, exclusion affects also the off-diagonal matrix entries, since by definition also haloes of different mass are distinct objects and can thus not overlap. Furthermore, different mass haloes are affected by non-linear clustering, since the probability to find any sort of massive object ($M>M_*$) in the vicinity of a massive object is enhanced. For simplicity we will employ equal number density mass bins, which all have the same fiducial shot noise $1/\bar{n}$.
\par
When trying to extract the amplitude and scale dependence of the noise, we need to remove all the contributions due to linear bias from the halo power spectra.
For this purpose, we will employ the stochasticity matrix as defined in \cite{Hamaus:2009op} (see also Eq.~\eqref{eq:snmatrixdiag} for the definition of the diagonal)
\be
(2\pi)^3 \ddir(\vec k+\vec k') C_{ij}(k)=\left\langle\left[\delta_i(\vec k)-b_{1,i} \delta(\vec k)\right]\left[\delta_j(\vec k')-b_{1,j} \delta(\vec k')\right]\right\rangle,
\label{eq:snmatrix}
\ee
such that we have in terms of the power spectra
\be
C_{ij}(k)=P_{ij}(k)-b_{1,i} P_{\delta j}(k)-b_{1,j} P_{\delta i}(k)+b_{1,i} b_{1,j} P_{\delta \delta}(k).
\label{eq:snmatrixpower}
\ee
The model introduced above can be straightforwardly generalized to multiple mass bins and their respective cross power spectra by the following replacements $b_1^2\to b_{1,i}b_{1,j}$, $b_2^2\to b_{2,i}b_{2,j}$ and $R\to R_{ij}=(R_i+R_j)/2$. The exact form of the combined exclusion radius is somewhat debatable, but for now we will employ the arithmetic mean. The resulting correction to the linear local bias model is given by
\begin{equation}
C_{ij}(k)=\frac{1}{\bar{n}}\delta^\text{(K)}_{ij}+\frac{1}{2}b_{2,i}b_{2,j} I_{22}(k,R_{ij})-V_{\text{excl},ij}W_{R_{ij}}(k)-b_{1,i}b_{1,j}V_{\text{excl},ij}[W_{R_{ij}}*P_\text{lin}](k),
\label{eq:sncorrmod}
\end{equation}
where $V_{ij}=4\pi/3 R_{ij}^3$. We see that the definition of the stochasticity matrix removes all occurrences of the linearly biased power spectrum. In Eulerian space both $P_\text{hm}$ and $P_\text{hh}$ would have an additional contribution from $b_2 I_{12}$, where $I_{12}$ describes the cross correlation between non-linear bias and non-linear matter clustering. The term is defined as
\be
I_{12}(k)=\int \dqc P_\text{lin,0}(q)P_\text{lin,0}(\left|\vec k-\vec q\right|)F_2(\vec q, \vec k-\vec q),
\ee
where $F_2(\vec q_1,\vec q_2)$ is the standard perturbation mode coupling kernel \cite{Bernardeau2002}.
The definition of the stochasticity matrix also removes all occurrences of $I_{12}$.
\section{Evaluations and Comparison to Simulations}\label{sec:simul}
\subsection{The Simulations \& Halo Sample}
Our numerical results are based on the Z\"urich horizon zHORIZON simulations, a suite of 30 pure dissipationless dark matter simulations of the $\Lambda$CDM cosmology in which the matter density field is sampled by $N_\text{p} = 750^3$ dark matter particles. The box length of $1500 \hMpc$, together with the WMAP3 \cite{Spergel:2007th} inspired cosmological parameters ($\Omega_\text{m}=0.25,\Omega_\Lambda=0.75,n_\text{s}=1,\sigma_8=0.8$), then implies a particle mass of $M_\text{p} = 5.55\tim{11}\hMs$. The total simulation volume is $V\approx 100 \hGpcc$ and enables precision studies of the clustering statistics on scales up to a few hundred comoving megaparsecs.
\par
The simulations were carried out on the ZBOX2 and ZBOX3 computer-clusters of the Institute for Theoretical Physics at the University of Zurich using the publicly available GADGET-II code \cite{Springel:2005mi}. The force softening length of the simulations used for this work was set to $60\ h^{-1}\text{kpc}$, consequently limiting our considerations to larger scales. The transfer function at redshift $z_\text{f} = 0$ was calculated using the CMBFAST code of \cite{Seljak:1996cmb} and then rescaled to the initial redshift $z_\text{i} = 49$ using the linear growth factor. For each simulation, a realization of the power spectrum and the corresponding gravitational potential were calculated. Particles were then placed on a Cartesian grid of spacing $\Delta x = 2 \hMpc$ and displaced according to a second order Lagrangian perturbation theory. The displacements and initial conditions were computed with the 2LPT code of \cite{Crocce:2006tr}, which leads to slightly non-Gaussian initial conditions.
 \par
Gravitationally bound structures are identified at redshift $z_\text{f}=0$ using the B-FoF algorithm kindly provided by Volker Springel with a linking length of $0.2$ mean inter-particle spacings. Haloes with less than 20 particles were rejected such that we resolve haloes with $M > 1.2\tim{13} \hMs$. The halo particles are then traced back to the initial conditions at $z_\text{i}=49$ and the corresponding centre of mass is identified. We split the halo sample into $10$ equal number density bins with a number density of $\bar n=3.72\tim{-5} h^{3}\text{Mpc}^{-3}$ leading to a shot noise contribution to the power at the level of $P_\text{SN}\approx2.7\tim{4} \hMpcc$.

\subsection{The Correlation Function in Lagrangian Space}\label{sec:correlsn}
\begin{figure}[t]
\centering
\includegraphics[width=0.49\textwidth]{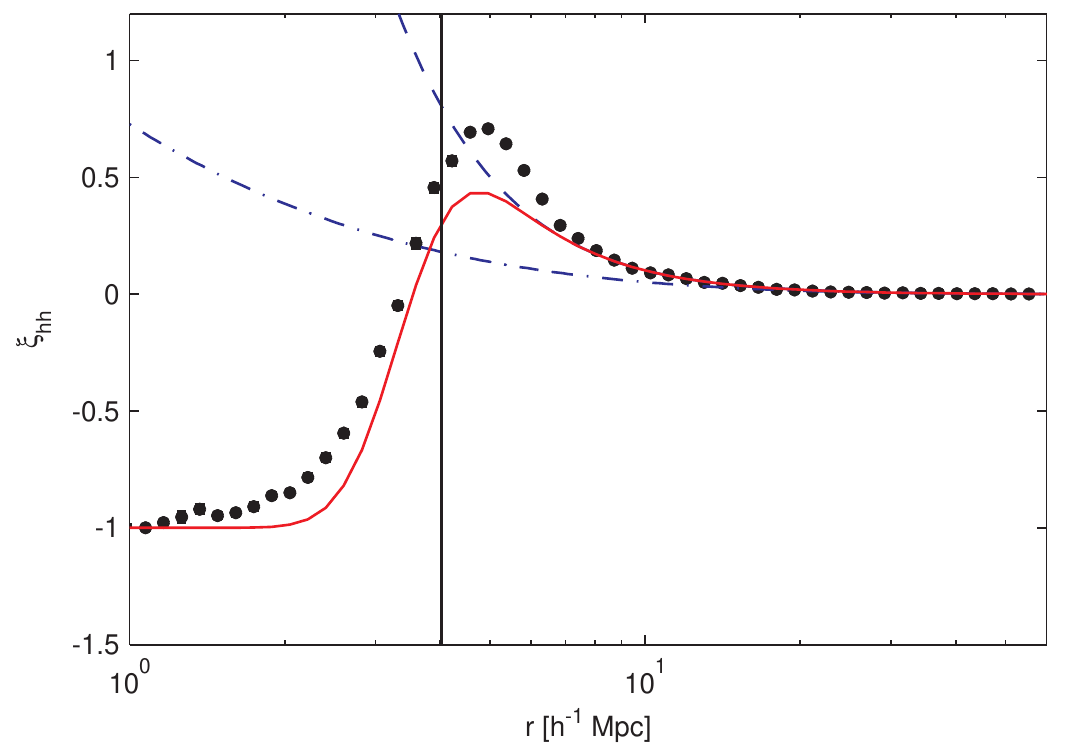}
\includegraphics[width=0.49\textwidth]{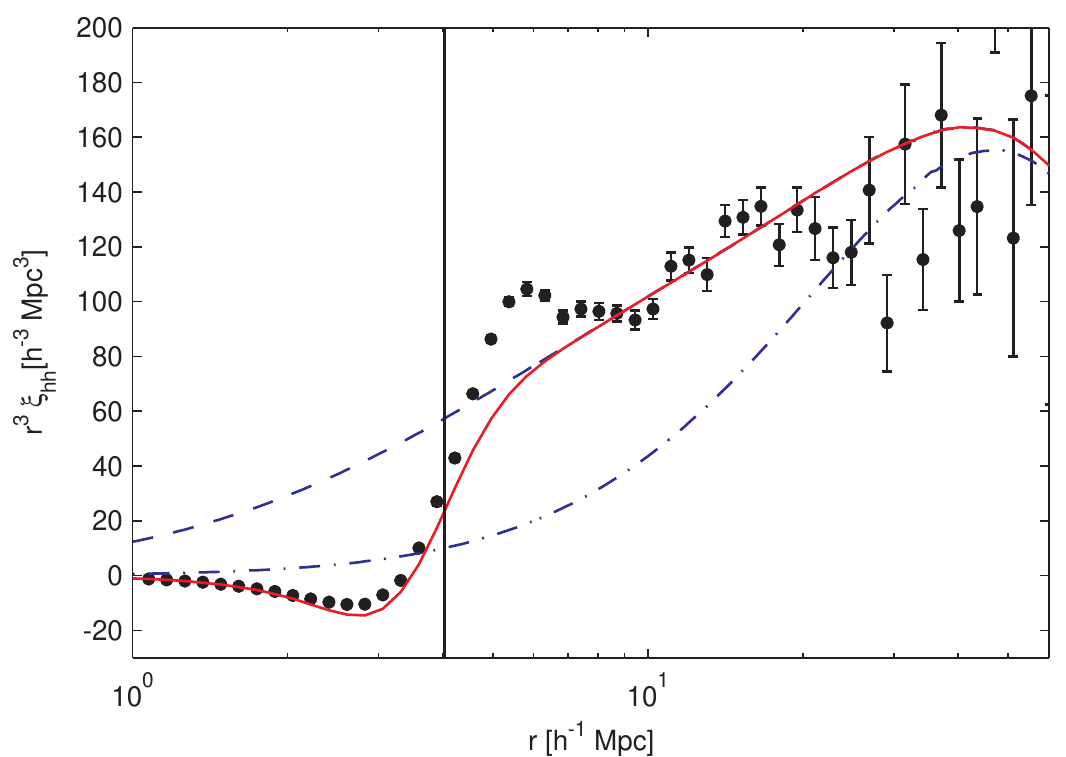}\\
\caption{Example of the halo-halo correlation function of the traced back haloes for mass bin V. The vertical solid line is the fitted exclusion radius. 
The dot-dashed line shows the linear bias contribution, whereas the dashed line shows linear plus quadratic bias. Note that the second order bias parameter was fitted to the correlation function and does deviate quite strongly from the PBS prediction.
The red solid line shows a simple model for halo exclusion Eq.~\eqref{eq:smoothedstep}. In the right panel we show the integrand of the Fourier transform $r^3 \xi_\text{hh}(r)$, which is of essential importance for the stochasticity modelling.}
\label{fig:correl_ic}
\end{figure}
\begin{figure}[t]
\centering
\includegraphics[width=0.49\textwidth]{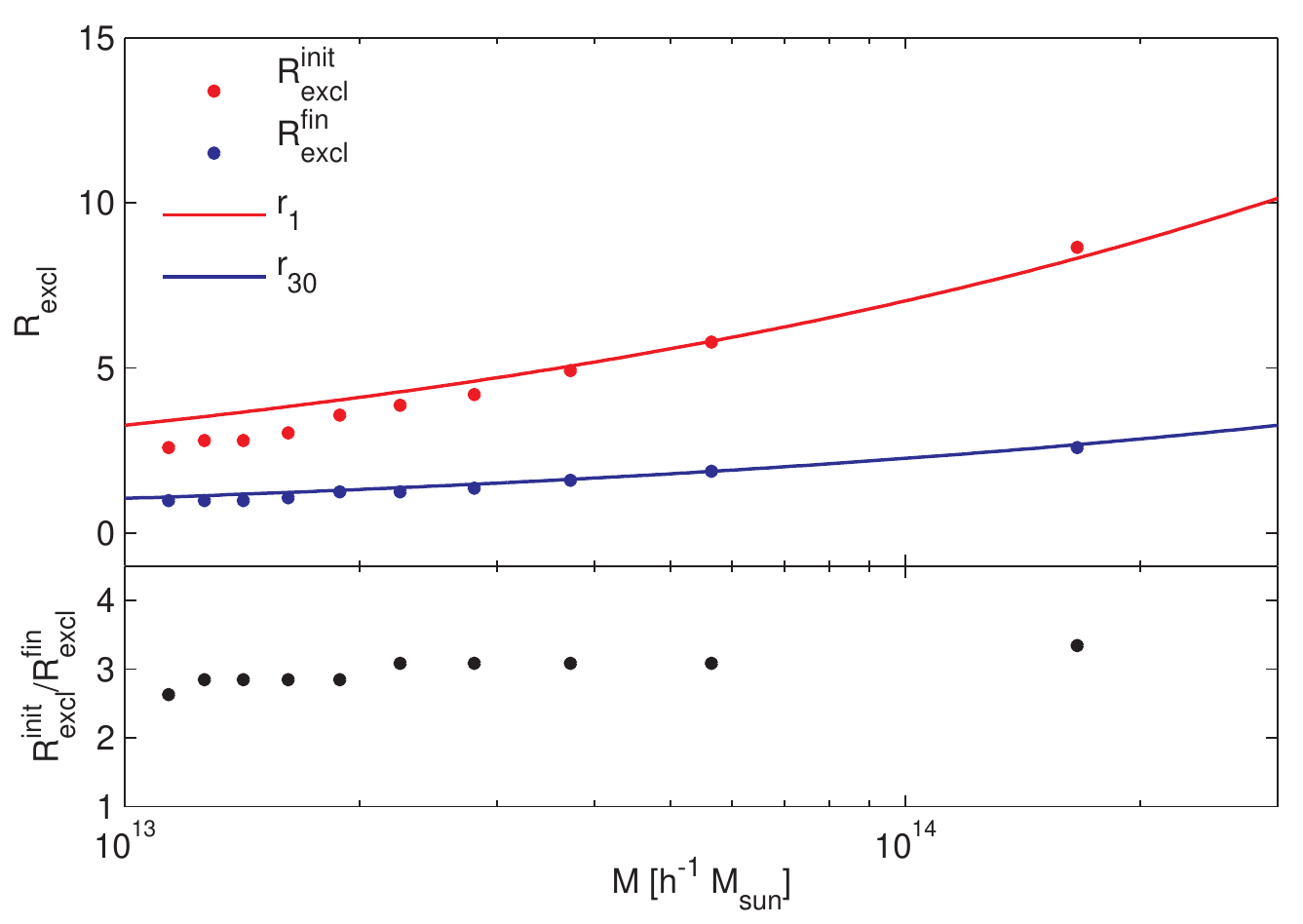}
\includegraphics[width=0.49\textwidth]{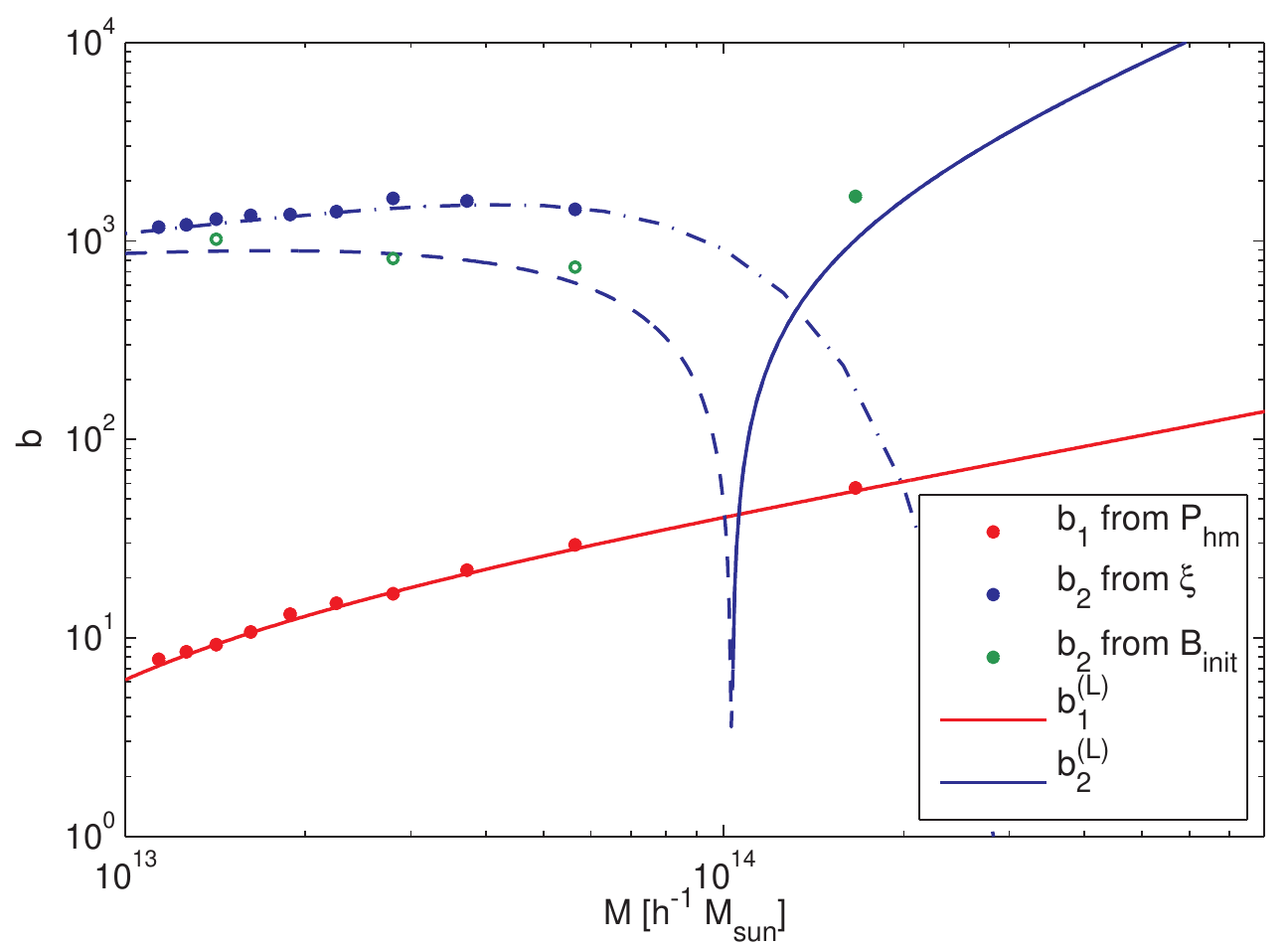}
\caption{\emph{Upper left panel: }Exclusion radii measured in the initial and final halo-halo correlation functions. The exclusion radius is defined as a fixed fraction of the maximum in the halo-halo correlation function (see Fig.~\ref{fig:correl_ic}). The red line shows the na\"{\i}ve estimate of the exclusion radius $R=(3M/4\pi \bar \rho)^{1/3}$, whereas the blue line shows $R=(3M/4\pi \bar \rho (1+\delta))^{1/3}$ with $\delta\approx 30$. This fitted overdensity is clearly distinct from the spherical collapse prediction of $\delta=180$. \emph{Lower left panel: }Ratio of the initial and final exclusion radii.
\emph{Right panel: }Bias parameters fitted from the halo-matter cross power spectrum ($b_1$), the halo-matter bispectrum and the correlation function ($b_2$). Note that $b_2$ fitted from the correlation function deviates strongly from the value predicted by the peak-background split (blue solid and dashed for positive and negative) and the value inferred from the bispectrum. The dash-dotted lines show a fit used for extrapolation purposes.}
\label{fig:exclrad}
\label{fig:biasparameters}
\end{figure}
The corrections to the halo power spectrum in our model are motivated by certain features in the halo-halo correlation function. While the fiducial stochasticity affects the correlation function only at the origin, the two other effects, exclusion and non-linear clustering, should be clearly visible in the correlation function at finite distances.
For this purpose we measure the correlation function of the traced back haloes for our $10$ halo mass bins using direct pair counting.
\par
In Fig.~\ref{fig:correl_ic} we show the correlation function for mass bin V. The log-linear plot clearly shows that the correlation function is $-1$ on small scales and shows a smooth transition to positive values around the exclusion scale visualized by the vertical black line. The exclusion scale is fitted both in the initial and final conditions as 0.8 times the maximum in the correlation function and is shown in Fig.~\ref{fig:exclrad}. The ratio between the initial and final exclusion radii is roughly $3$ for all mass bins. The spherical collapse model suggests that haloes collapse by a factor $5$, but there is no reason to believe that protohaloes that are in direct contact in Lagrangian space are still touching each other in Eulerian space. Thus it is reasonable to expect a somewhat smaller reduction in the exclusion scale.
On large scales $30 \hMpc<r<90 \hMpc$ the correlation function is reasonably well described by linear bias shown in the Figure as a dot-dashed line.
We infer the linear bias parameter from the ratio of halo-matter cross power spectrum and matter power spectrum on scales $k< 1.5 \tim{-2} \ihMpc$
\be
\hat{b}_\text{1,hm}=\frac{\hat{P}_\text{hm}}{\hat{P}_\text{mm}}.
\ee
See Fig.~\ref{fig:scaledepbiasinit} and App.~\ref{app:peakeffects} for why we have to restrict the linear bias fitting to large scales even in the initial conditions.
The advantage of the cross power spectrum is that it should be free of stochasticity contributions and fully described by linear bias \cite{Frusciante:2012la} on large scales.\\
There is a clear enhancement of the data in Fig.~\ref{fig:correl_ic} compared to the linear bias model on small scales. Thus we consider the quadratic bias model
\be
\xi_\text{hh}^\text{(c)}(r)=b_1^2 \xi_\text{mm,lin}(r)+\frac{1}{2}b_2^2 \xi_\text{mm,lin}^2(r)
\label{eq:xibiassecond}
\ee
and fit for the quadratic bias parameter on scales exceeding the maximum. The resulting continuous correlation function is shown as the dashed line in Fig.~\ref{fig:correl_ic}. It does not fully account for the enhanced clustering outside the exclusion scale. 
The inferred bias parameters are shown in Fig.~\ref{fig:biasparameters} and will be discussed in more detail below. Note that the above fit is performed using an un-smoothed version of the linear correlation function, whereas the local bias model relies on an explicit smoothing scale. Here, we argue that the smoothing scale for the local bias model should be related to the Lagrangian scale of the haloes and thus be smaller than the exclusion scale. In this case, the smoothing scale does typically not affect the scale dependence of the correlation function (except for around the BAO scale).

In our above discussion we have assumed a sharp transition between the exclusion regime and the clustering regime. This is certainly unphysical as is obvious in Fig.~\ref{fig:correl_ic}.  The smoothness is probably caused by a number of phenomena, for instance mass variation within the mass bins (should be quite small $<1.3 \hMpc$ for the highest mass bin which has $R\approx 8 \hMpc$ and even smaller for the lower mass bins) or alignment of triaxial haloes.
The lack of a physically motivated, working model for the transition forces us to employ a somewhat ad hoc functional form for the step, which is based on a lognormal distribution of halo distances (see App.~\ref{app:exclkern})
\be
\xi^\text{(d)}_\text{hh}(r)=\frac{1}{2}\left[1+\text{erf}\left(\frac{\log_{10}(r/R)}{\sqrt{2}\sigma}\right)\right]\bigl[\xi^\text{(c)}_\text{hh}(r)+1\bigr]-1.
\label{eq:smoothedstep}
\ee
For alternative implementations of exclusion windows in the context of the halo model see \cite{Tinker:2005on,Valageas:2011co}.
The resulting shape of the correlation function is shown as the red solid line, where the smoothing was chosen to be $\sigma\approx 0.09$ and seems to be quite independent of halo mass. The model clearly underestimates the peak in the data in the log-linear plot.
Our final goal is to construct an accurate model for the effect of exclusion and non-linear clustering on the power spectrum. Thus, we should not only check the validity of our model on plots of the correlation function itself but also on the integrand in the Fourier integrals $r^3 \xi_\text{hh}(r)\derd \ln r$. We do so in the right panel of Fig.~\ref{fig:correl_ic}, where it is obvious that the model does not reproduce the exact shape of the correlation function outside of the exclusion scale. However, we certainly improved over the naive linear biasing on all scales and obtained a reasonable parametrization of exclusion and non-linear clustering effects.
\par
Let us now come back to the mass dependence of the inferred bias and exclusion parameters. As we show in Fig.~\ref{fig:biasparameters}, the linear bias $b_1$ is in very good agreement with the bias parameters inferred from a Sheth-Tormen mass function \cite{Scoccimarro:2001ho} rescaled to the initial conditions at $z_\text{i}=49$. For the second order bias we compare the measurement from the correlation function to measurements from the bispectrum of the protohaloes and the second order bias inferred from the Sheth-Tormen mass function. The latter agrees reasonably well with the bispectrum measurement reproducing the zero crossing in the theoretical bias function. Note that the bispectrum measurement (for details of the approach see \cite{Baldauf:2012ev}) uses only large scale information and is thus a clean probe for second order bias. Isolating second order bias effects in the correlation function is less straightforward. With decreasing scale higher and higher bias parameters become important, and to our knowledge there is no established scale down to which a certain order of bias can be trusted to a given precision. Our fitting procedure was led by the goal of obtaining a good parametrization of the correlation function, which could subsequently be used to calculate the corresponding power spectra and the corrections to the linear bias model. Note that due to the functional form of Eq.~\eqref{eq:xibiassecond} this fitting approach allows inference of the magnitude of $b_2$, but not its sign. The second order bias parameters obtained in this way deviate significantly from the theoretical bias function and the bispectrum measurement.  The most severe failure of the model is the imaginary $b_2$ for the highest mass bin. In this case the deviation is connected to corrections arising from the peak constraint, as we explain in App.~\ref{app:peakeffects}. We find that the initial Lagrangian second order bias parameter in Fig.~\ref{fig:biasparameters} can be roughly fitted as follows
\begin{align}
b_2=&1100\left(\frac{M}{10^{13} \hMs}\right)^{0.35}\eh{-\left(\frac{M}{10^{14}\hMs}\right)^2}.
\label{eq:b2Mrelation}
\end{align}
We will use this fitting function for extrapolation in mass and redshift in \S \ref{sec:rscorr}. The initial second order bias parameters for halo samples identified at higher redshifts ($z=0.5$ and $z=1$) are roughly the same as for the $z=0$ halo sample. \\
Although we will use this $b_2$ fit to predict the resulting stochasticity corrections, we do not believe that the observed scale dependence of the correlation function is solely a second order bias effect. We checked an expansion of Eq.~\eqref{eq:xibiassecond} to higher orders in the correlation function using peak-background split bias parameters. Even up to tenth order there is no considerable improvement in the fit. Thus we argue that the enhancement is a non-perturbative effect (e.g. peak bias) and consider the $\xi^2$ scale dependence as a reasonably well working phenomenological parametrization rather than a physical truth. We hope to shed more light on this issue in a forthcoming paper.
\subsection{Stochasticity Matrix}
\begin{figure}[t]
\includegraphics[width=0.49\textwidth]{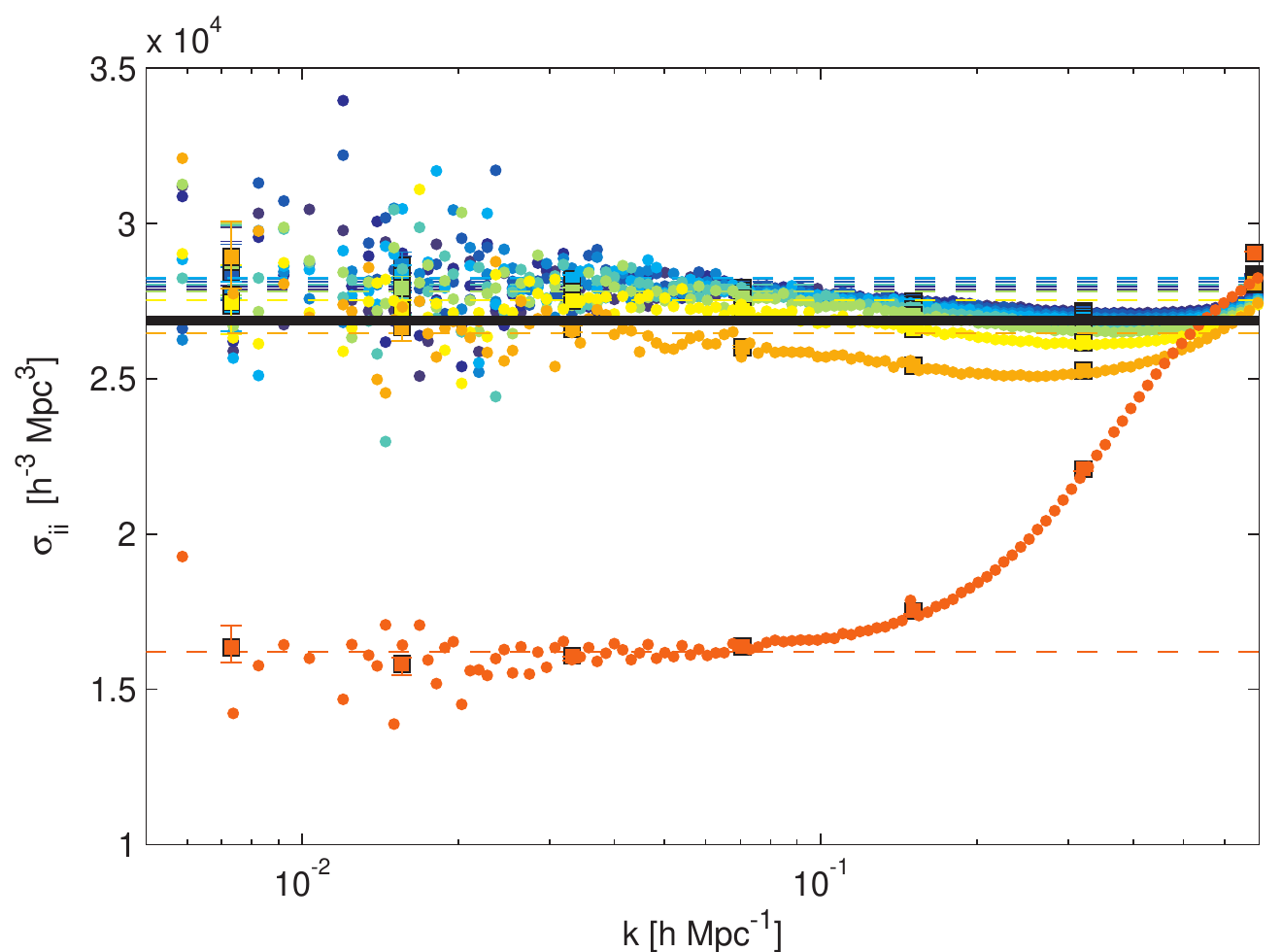}
\includegraphics[width=0.49\textwidth]{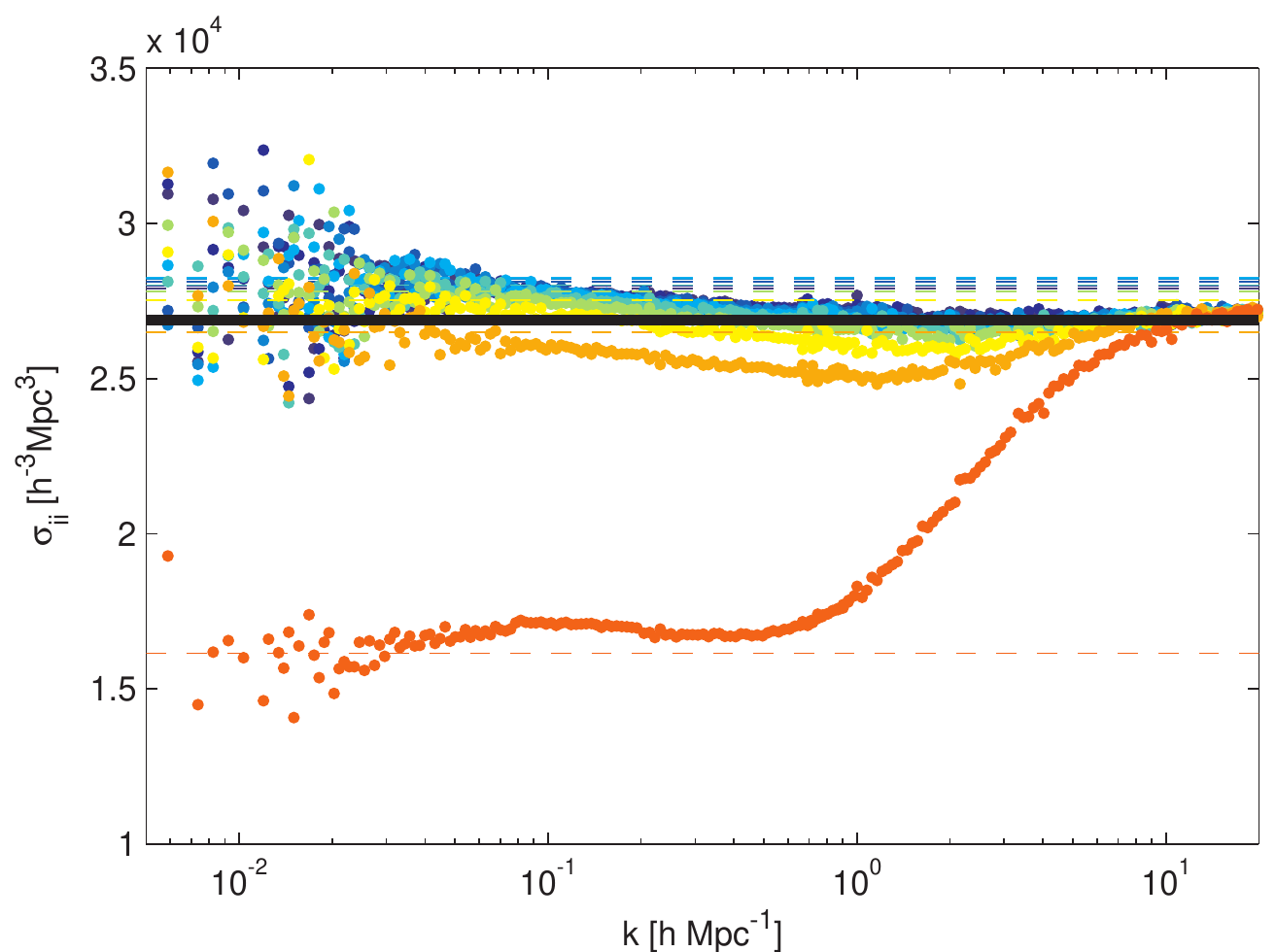}
\caption{Diagonals of the stochasticity matrix $\sigma_{ij}$. \emph{Left panel: }Initial conditions $z_\text{i}=49$ \emph{Right panel: }Final field $z_\text{f}=0$.
There is remarkable agreement in the large scale amplitude between initial conditions and final field besides the strong difference in the bias parameters and growth factors. We highlight this fact by the horizontal dashed lines that have the same amplitude in both panels and are matched to the large scale stochasticity matrix in the initial conditions.
For both panels, there is clear evidence for stochasticity going to fiducial $1/\bar{n}$ for high wavenumbers, and a modification due to exclusion and clustering for $k\leq 1/R$, where $R$ is the scale of the halo at the corresponding redshift. Note the different scaling of the $k$-scale in the two plots. The mass increases from blue to orange, i.e., top to bottom.}
\label{fig:snmatsim}
\end{figure}
In Fig.~\ref{fig:snmatsim} we show the diagonals of the stochasticity matrix measured in our simulations in Lagrangian space ($z_\text{i}=49$) and Eulerian space ($z_\text{f}=0$). The most remarkable observation in this plot is the agreement between the results, given the different amplitude of the growth factors and the linear bias parameters at these two times. This is a result of the fact, that gravity can not introduce or alter $k^0$ dependencies \cite{Peebles:1980th} due to mass and momentum conservation. This can for instance be seen in the low $k$-limit of standard perturbation theory \cite{Bernardeau2002}: The mode coupling term $P_{22}$ is a gravity-gravity correlator and thus scales as $k^4$, whereas the propagator term $P_{13}$ is a gravity-initial condition correlator and scales as $k^2 P_\text{lin}$.\\
For the highest mass bin there is a clear suppression of the noise level on the largest scales which then asymptotes to the fiducial value $1/\bar{n}_\text{h}$ at a scale $k\approx 1/R \approx 0.3 \ihMpc$. Since the radius of the halo shrinks during collapse, this scale is found at a higher wavenumber in Eulerian space. For the less massive haloes the behavior is not completely monotonic. On large scales we find a noise level slightly exceeding the fiducial value. Going to higher wavenumbers the fiducial value is crossed, the residual reaches a minimum and finally asymptotes to $1/\bar{n}$. This behavior can be explained as follows: the clustering scale exceeds the exclusion scale and as we have argued in \S \ref{ssec:toyclust}, the enhanced correlation on the clustering scale leads to a positive contribution on the largest scales that decays for lower wavenumbers than the negative exclusion correction.
\par
The wavenumber at which the stochasticity asymptotes to its fiducial value in the final field is at fairly high wavenumbers, exceeding the Nyquist frequencies of both $N_\text{c}=512$ and $N_\text{c}=1024$ grids. To probe smaller scales we employ a mapping technique \cite{Jenkins:1998ev,Smith:2003st} that allows us to resolve small scales without having to increase the grid size beyond $N_\text{c}=512$. The technique consists of splitting the box into $n$ parts per dimension and adding these parts to the same grid. This allows inference of each $l$-th mode but also increases the Nyquist frequency by a factor $l$. We use several different mapping factors $l=4,6,12,20,50$ to probe all the scales up to $k\approx 20 \ihMpc$.
\par
\begin{figure}[t]
\includegraphics[width=0.49\textwidth]{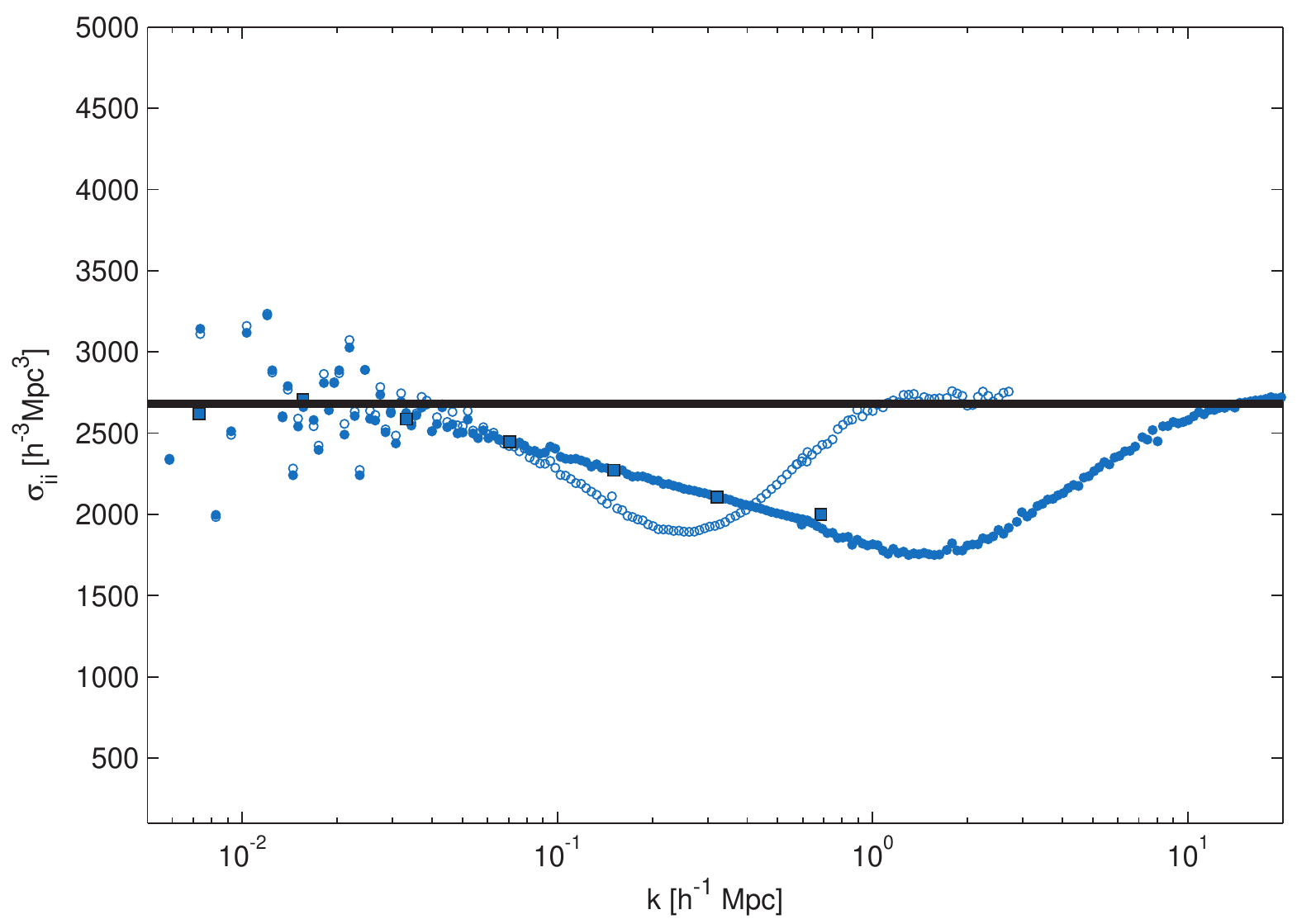}
\includegraphics[width=0.49\textwidth]{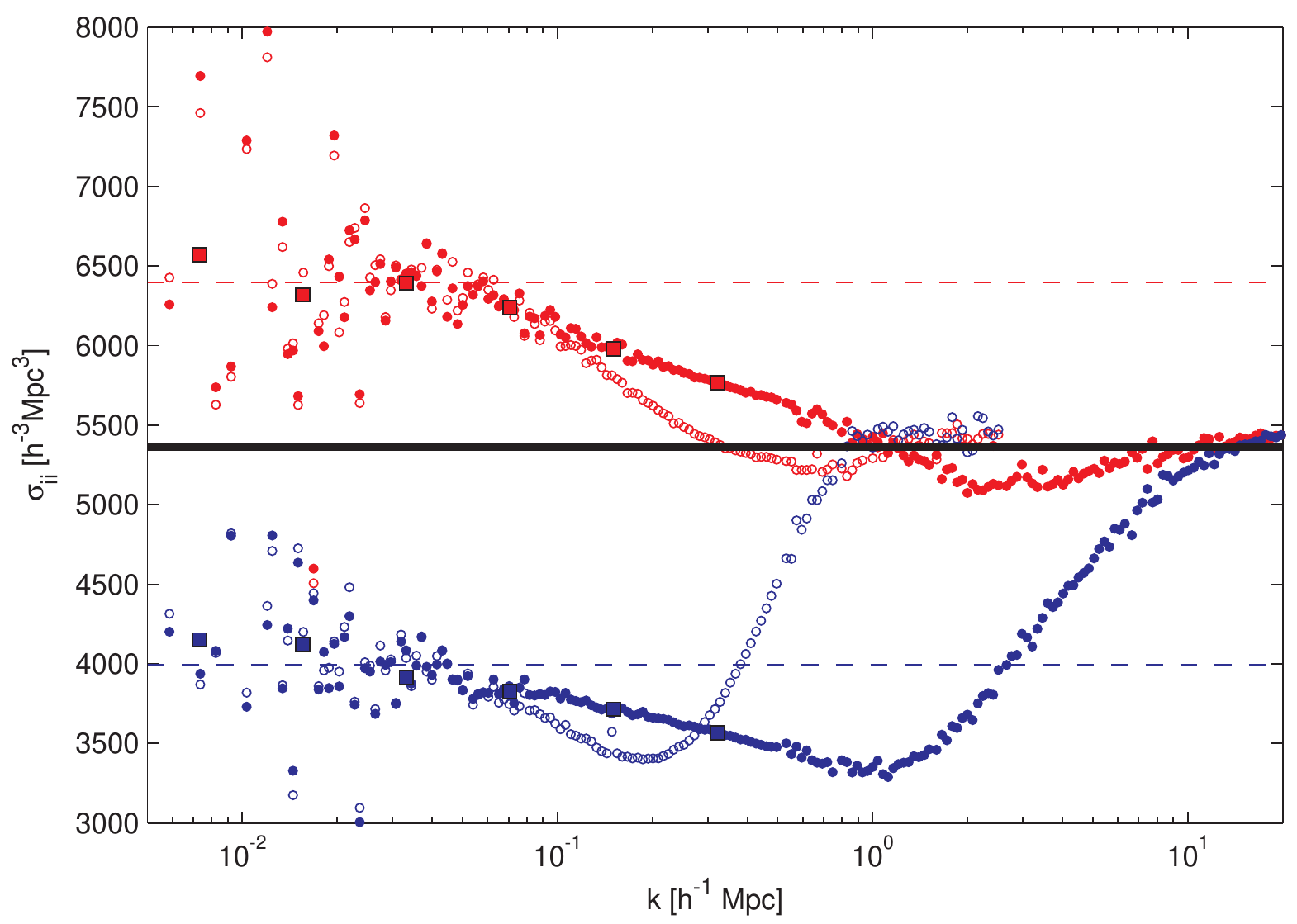}
\caption{\emph{Left panel: }Diagonals of the stochasticity matrix for one mass bin containing all the haloes in our simulation. Open points show the initial condition measurement, whereas the filled points show the final value. The horizontal thick solid line shows the fiducial shot noise $1/\bar{n}=2680 \hMpc$. The Eulerian bias is $b_1^\text{(E)}=1.49$. On the largest scales there seems to be a cancellation between the exclusion and non-linear clustering contributions resulting in no net correction to the fiducial shot noise. \emph{Right panel: }Same as left panel but for splitting our haloes into two mass bins with equal number density. The upper red points are the measurement for the lighter, lower bias bin $b_1^\text{(E)}=1.25$ and the lower blue points are for the more massive, higher bias bin $b_1^\text{(E)}=1.74$. The fiducial shot noise is $1/\bar{n}=5362 \hMpc$. Note that both mass bins show a significant scale dependence of the stochasticity.}
\label{fig:snmatsim12bins}
\end{figure}

In Fig.~\ref{fig:snmatsim12bins} we show the diagonals of the stochasticity matrix for one and two mass bins respectively. For the one bin case we select all the haloes in our simulation, effectively combining all the ten mass bins resulting in $M_\text{1bin}=3.84\tim{13}\hMs$. For the two bin case we combine the five lightest and the five heaviest mass bins resulting in masses $M_\text{2bin,I}=1.47\tim{13}\hMs$ and $M_\text{2bin,II}=6.21\tim{13}\hMs$. The plot shows both the initial condition and the final stochasticity and the two agree very well in the low-$k$ limit. The stochasticity correction does not depend on the fiducial stochasticity and thus the scale dependence of the total stochasticity is more pronounced in the wider mass bins due to their lower fiducial shot noise. Interestingly the stochasticity correction for the 1-bin case vanishes in the low-$k$ limit, but is negative in the intermediate regime. This is a sign of a perfect cancellation between exclusion and non-linear clustering. The final stochasticity seems to be a $k$-rescaled version of the initial stochasticity. Finally, let us stress that the power spectrum of the wide bins can not be obtained by a summation of the contributing bin-power spectra from the ten bin case since one has to account for the off-diagonal components of the stochasticity matrix.
\par
\begin{figure}[t]
\includegraphics[width=0.49\textwidth]{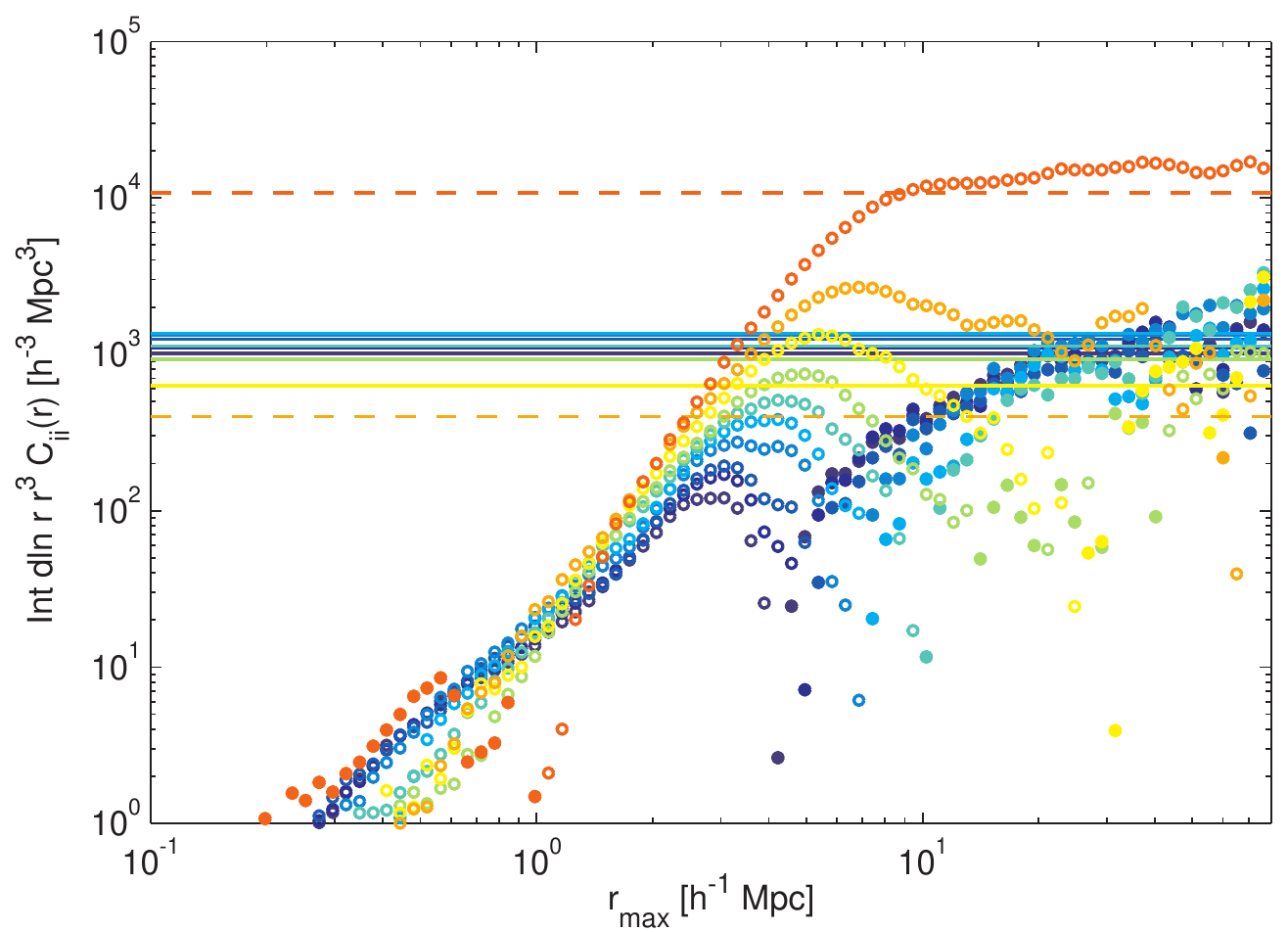}
\includegraphics[width=0.49\textwidth]{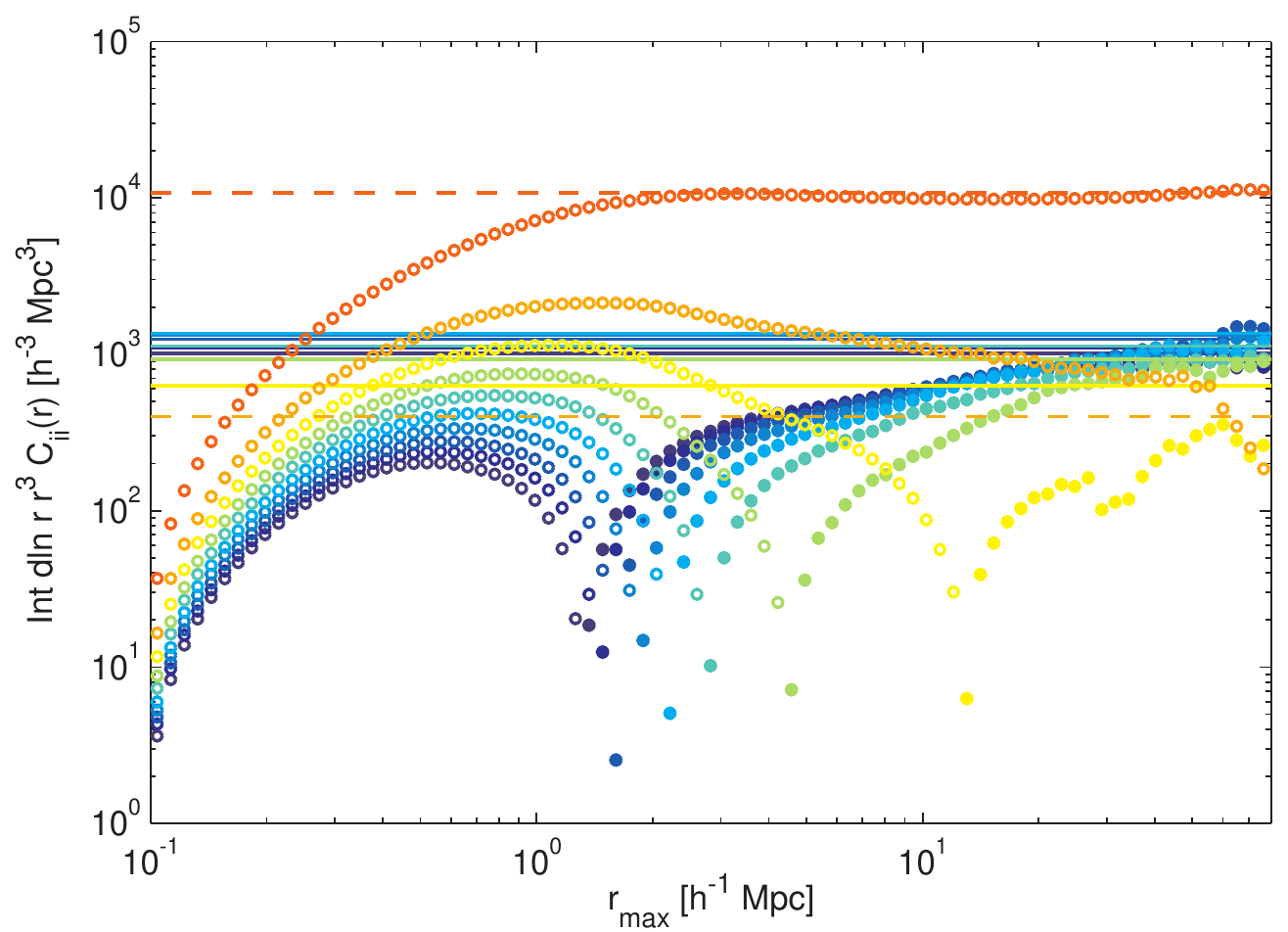}
\caption{Cumulative contributions to the stochasticity correction up to scale $r$ (see Eq.~\ref{eq:sncontrib}) in configuration space for our 10 mass bin sample. Open symbols show negative contributions and filled symbols positive contributions. \emph{Left panel: }Initial conditions. \emph{Right panel: }Final configuration at $z=0$. We clearly see, that the corrections in the initial conditions are dominated on larger scales than in the final configuration, where the negative corrections are clearly in the one halo regime, where both the halo-matter and matter-matter correlation functions are highly non-linear. This gives further motivation for the modelling of the effect in Lagrangian space.
The horizontal solid (dashed) lines show the positive (negative) stochasticity corrections inferred from Fig.~\ref{fig:snmatsim}.}
\label{fig:snmatreal}
\end{figure}

In the left panel of Fig.~\ref{fig:biascorr} we compare our model Eq.~\eqref{eq:sncorrmod} to the measured stochasticity matrix of the ten bins in the initial conditions. The only modification to the model is that we replace the hard cutoff by the smoothed transition  Eq.~\eqref{eq:smoothedstep}. We employ the parameters obtained in the fit to the corresponding halo correlation functions in \S \ref{sec:correlsn}. The data points in the plot are copied from Fig.~\ref{fig:snmatsim}. Given the differences between the model and the correlation functions in Fig.~\ref{fig:correl_ic}, there is a reasonably good agreement both in large scale amplitude and scale dependence of the stochasticity correction. We can conclude that while being a relatively crude fit to the correlation function, our model can account for the major effects, exclusion and non-linear clustering. The drawback is that this model lives in Lagrangian space and can not be straightforwardly applied to the halo power spectrum in Eulerian space.\\
In Fig.~\ref{fig:crossterms} we show the off-diagonal terms of the stochasticity matrix in the initial conditions and our corresponding model predictions. As for the diagonals discussed above, the corrections can be either positive or negative, depending on whether exclusion or nonlinear clustering dominates The model predictions are in reasonable agreement with the measurements except for the highest mass bin. This failure is connected to the fact that the $b_2$ parameter for the highest mass bin is imaginary, i.e., we have to set the second order bias term in the cross correlations to zero. This is a severe problem of our overly simplistic model, which is related to the importance of the peak effects for the correlation function of the highest mass bin (see App.~\ref{app:peakeffects}).\\
Let us try to gain some more insight on where the stochasticity corrections arise in Lagrangian and Eulerian space. For this purpose we consider the configuration space version of the diagonal of the stochasticity matrix defined in Eq.~\eqref{eq:snmatrixpower}
\be
C_{ii}(r)=\xi_{ii}(r)-2b_{1,i}\xi_{i\delta}(r)+b_{1,i}^2\xi_{\delta\delta}(r).
\ee
The stochasticity level in the $k\to 0$ limit is then given by
\be
C_{ij}(k)\xrightarrow{k\to0}\int_0^\infty \derd \ln r \ r^3 \sigma_{ij}(r).
\label{eq:sncontrib}
\ee
In Fig.~\ref{fig:snmatreal}, we show the above integral as a function of the upper integration boundary where the full large scale stochasticity correction would be obtained by taking this boundary to infinity. Comparing the contributions in the initial conditions and the final configuration, we clearly see that the large scale stochasticity arises on different scales at the two times. We clearly see that the negative stochasticity corrections are dominated at much smaller scales in the final configuration as compared to the initial conditions. At these scales the halo-matter and matter-matter correlation functions are dominated by the one halo term, i.e., the halo profile, which complicates quantitative predictions of the exclusion effect in the final configuration and motivates our Lagrangian approach.

\subsection{Redshift and Mass Dependence of the Correction}\label{sec:rscorr}
\begin{figure}[t]
\includegraphics[width=0.49\textwidth]{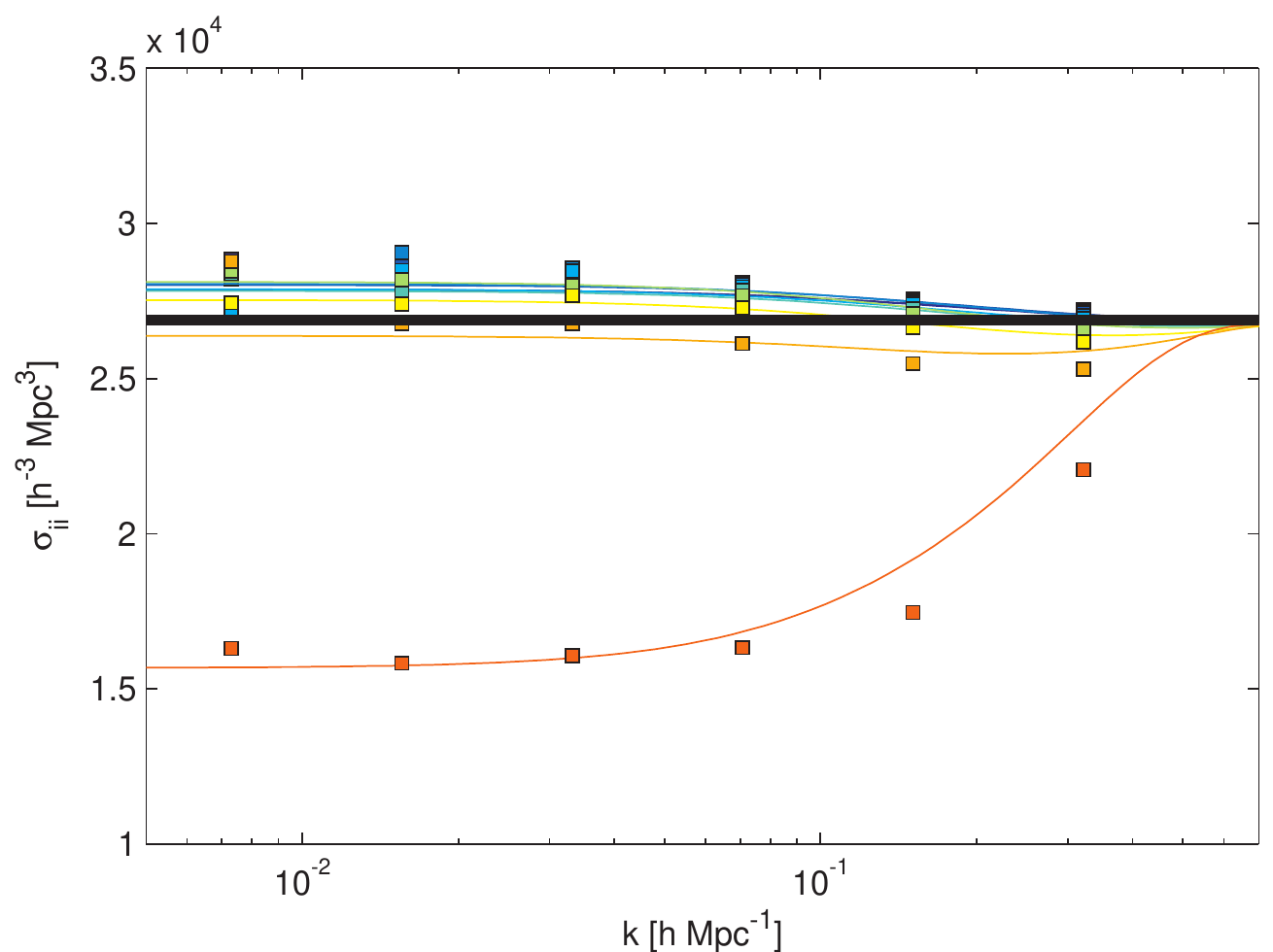}
\includegraphics[width=0.49\textwidth]{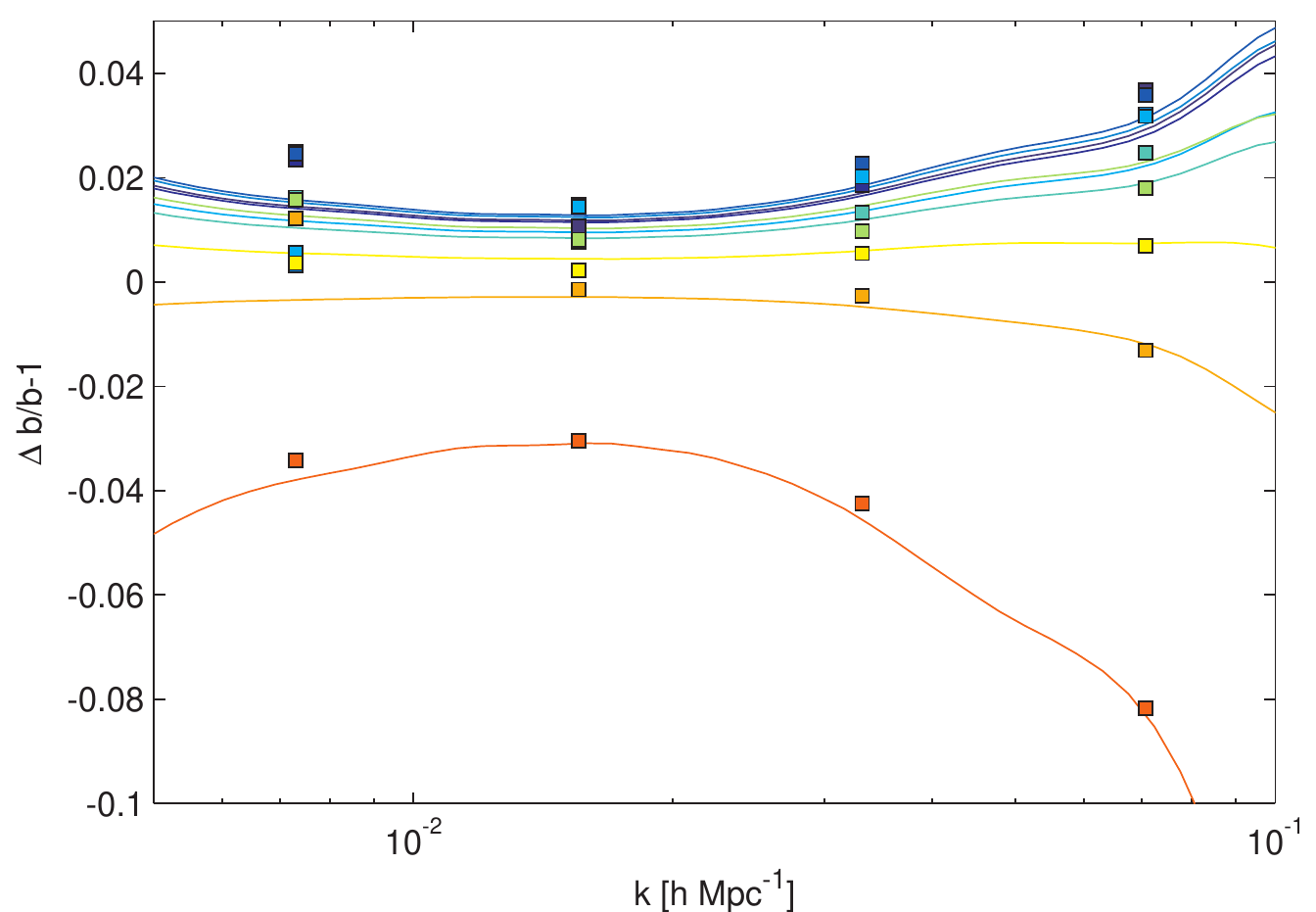}
\caption{\emph{Left panel: }Theoretical prediction for the diagonals of the stochasticity matrix for the ten mass bins. The parameters $b_2$ and $R$ are measured from the correlation function. The points are the coarsly binned simulation measurement taken from the left panel of Fig.~\ref{fig:snmatsim}.
\emph{Right panel: }Bias correction as defined in Eq.~\eqref{eq:fracbiascorr}. While there is clear scale dependence for weakly non-linear wavenumbers $k> 0.1 \ihMpc$ the correction is fairly flat on large scales, where the cosmic variance errors are largest. It could thus be easily interpreted as a higher or lower bias value. The points show the relative deviation of the bias inferred from the halo-halo power spectrum in the final configuration from the bias inferred from the halo-matter cross power spectrum.}
\label{fig:biascorr}
\end{figure}
The subtraction of the fiducial $1/\bar{n}$ shot noise from the power spectrum will lead to a biased estimate of the continuous halo power spectrum. Thus, estimating the bias as 
\be
\hat{b}_\text{1,hh}=\sqrt{\frac{\hat{P}_\text{hh}-1/\bar{n}}{\hat{P}_\text{mm}}}
\ee
will lead to a flawed estimate of the bias. Indeed, it has been found in simulations that the biases estimated from the auto- and cross-power spectra are generally not in agreement \cite{Elia:2011th,Okumura:2012rs}.
Studying for instance Table I in \cite{Okumura:2012rs} we see that $\hat{b}_\text{1,hh}$ exceeds $\hat{b}_\text{1,hm}$ for low mass haloes at redshift $0$ indicating that the fiducial shot noise subtraction underestimates the true noise level. For high mass objects the opposite happens, the bias from the cross-power exceeds the bias from the auto power indicating that the employed $1/\bar{n}$ shot noise subtraction overestimates the true noise level.\\
Let us try to understand this effect in more detail. Based on our model, subtraction of the fiducial shot noise on large scales leaves us with the linear bias term plus the stochasticity correction
\be
\hat{P}_\text{hh}(k)-\frac{1}{\bar{n}}=\Delta P_\text{hh}(k)+b_1^2 P_\text{lin}(k)=\hat{b}_\text{1,hh}^2 \hat{P}_\text{mm}(k),
\ee
where in absence of shot noise in the cross-power spectrum $b_1=\hat{b}_\text{1,hm}$.
Thus we have for systematic error on the linear bias parameter
\be
\frac{\Delta b_1}{b_1}=\frac{\hat{b}_{1,\text{hh}}}{b_1}-1\approx \frac12 \frac{\Delta P_\text{hh} }{b_1^2 P_\text{mm}}\label{eq:fracbiascorr}
\ee
Consequently the ratio $\hat{b}_\text{1,hh}/\hat{b}_\text{1,hm}$ is a function of mass and redshift due to the mass and redshift dependence of the parameters of the model. We show the $k$-dependence of the bias correction in Fig.~\ref{fig:biascorr}. Since we don't have a reliable model to relate the scale dependent stochasticity matrix from Lagrangian space to the one in Eulerian space we employ the scale dependent stochasticity model from Lagrangian space but divide by the present day linear power spectrum. This procedure should provide a reasonable estimate for the bias corrections on large scales. The linear bias is usually estimated close to the peak of the linear power spectrum, where it is approximately flat and where non-linear corrections are believed to be negligible. As a result, the bias correction is also fairly flat and would lead to a $1\%$ overestimation of bias for low mass objects and a $3-4\%$ underestimation for clusters. This behaviour can qualitatively explain the deviations found in \cite{Okumura:2012rs}.
\par
\begin{figure}[t]
\centering
\includegraphics[width=0.49\textwidth]{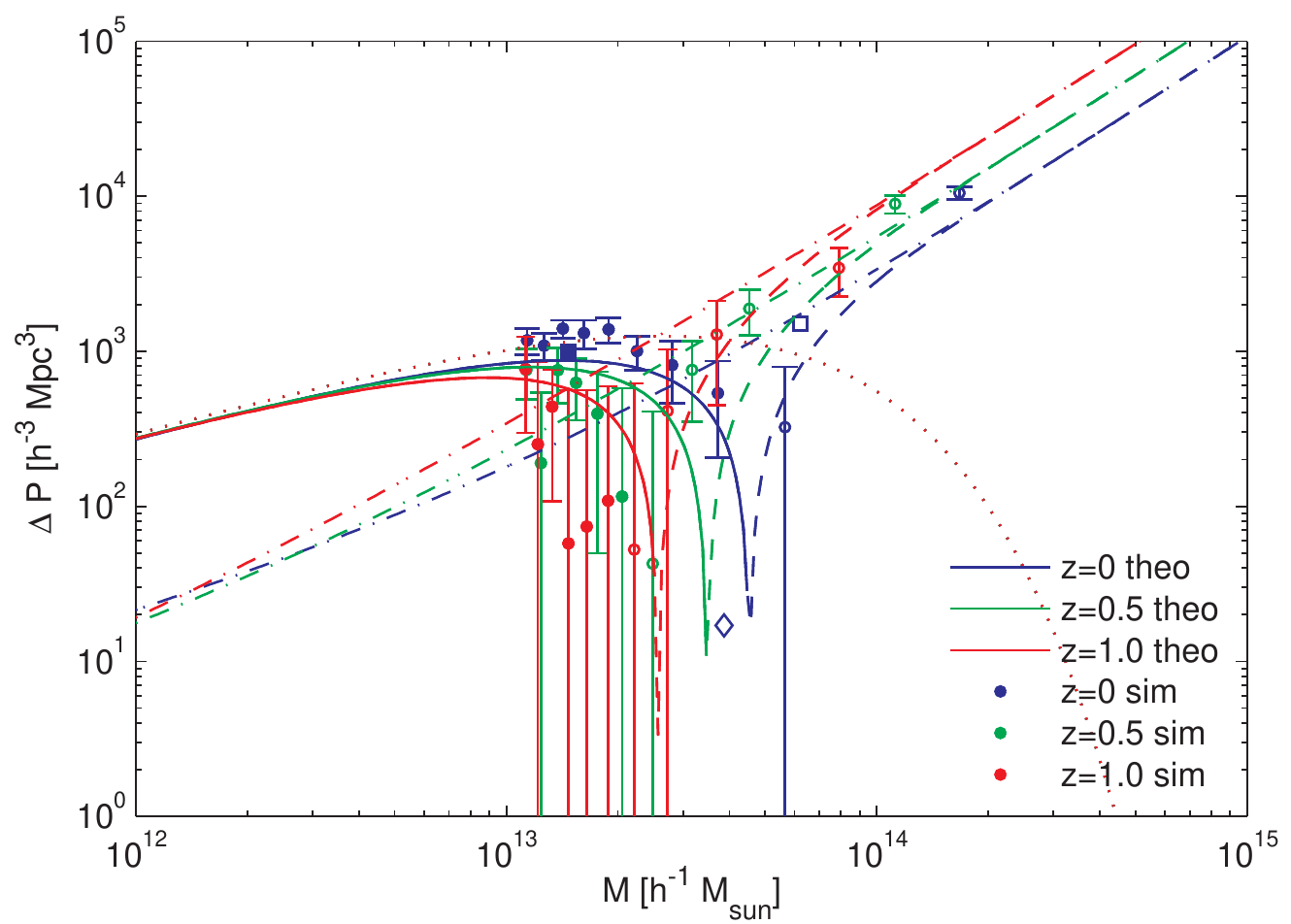}
\caption{Dependence of the stochasticity correction on halo mass and redshift. The lines are based on linear bias parameters from the peak-background split and a fitting function modelling $b_2$. The dashed lines and open points describe negative values. We also include the low-$k$ limits of the one and two bin splitting as the diamond and squares, respectively. The dotted line shows the positive low mass correction arising from non-linear biasing whereas the dash-dotted blue, green and red lines show the negative high mass exclusion corrections for the three different redshifts.}
\label{fig:sncorrmassdep}
\end{figure}

In Fig.~\ref{fig:sncorrmassdep} we show the amplitude of the low-$k$ limit of the stochasticity correction for ten equal halo mass bins at redshifts $z=0,0.5,1$. We overplot the theoretical expectation based on our model, linear bias parameters from the peak background split and second order bias parameters obtained from our phenomenological $b_2$ relation in Eq.~\eqref{eq:b2Mrelation}. In particular, we calculate the Lagrangian bias parameters and exclusion radii corresponding to the halo samples at $z=0,0.5,1$ and use them to predict the stochasticity correction. As a general result we can see that there is a negative correction for high masses and a positive correction for low masses with a zero crossing scale that decreases with increasing redshift. The model captures the trends in the measurements relatively well. We are also overplotting the low-$k$ amplitude of the SN for the one and two bin samples at $z=0$ as the squares and diamonds. Besides the fact that these bins are much wider and thus have lower fiducial shot noise, the low-$k$ amplitude is in accordance with the model and also narrower mass bins of the same mass. This fact supports our conjecture, that the stochasticity correction does not depend on the fiducial shot noise, but rather on mass (via the exclusion scale and the linear and non-linear bias parameters).
The negative correction for high masses is dominated by the exclusion term, whose amplitude depends on the linear bias parameter and the exclusion scale. The latter is a function of mass but not a function of redshift, whereas the bias increases with redshift and thus the negative correction at high masses also increases with redshift. The positive correction at the low mass end depends on the second order bias parameter. In our fits to the correlation function we found that this parameter is roughly constant for the three different redshifts under consideration.
\subsection{Eigensystem and Combination of Mass Bins}
\begin{figure}[t]
\centering
\includegraphics[width=0.49\textwidth]{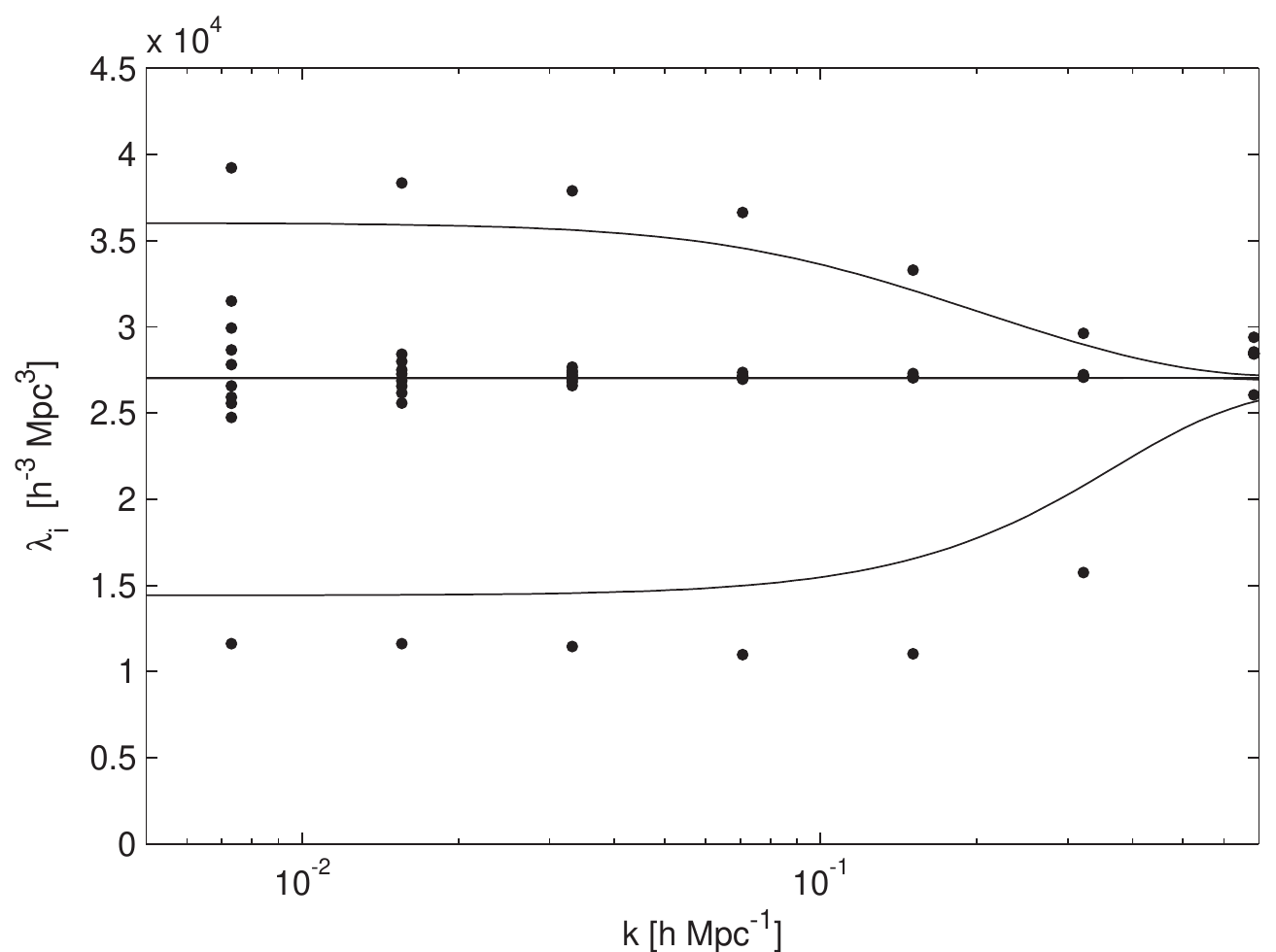}
\includegraphics[width=0.49\textwidth]{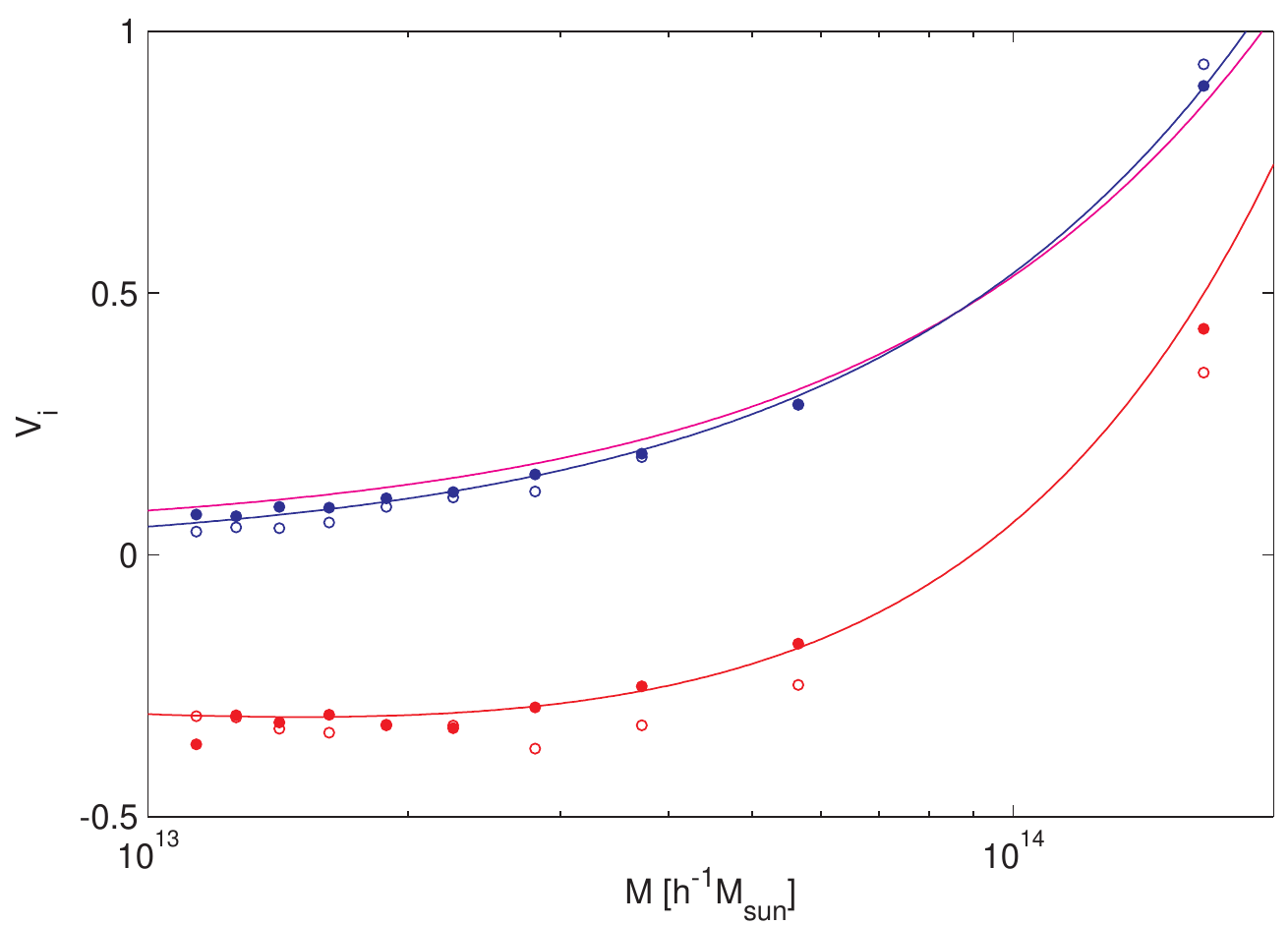}
\caption{ \emph{Left panel: }Eigenvalues of the stochasticity matrix for the traced back $z=0$ haloes. The lines show our prediction based on the modelling of the initial correlation function whereas the points are based on the diagonalization of the measured stochasticity matrix of the traced back haloes. \emph{Right panel: }Eigenvectors in the $k\to 0$ limit. The black and red filled points show the measured eigenvectors for the lowest and highest eigenvalue in the initial conditions, whereas the open points show the respective prediction of our model. The blue line shows mass weighting, the red line $b_2$-weighting and the magenta line shows the modified mass weighting proposed by \cite{Hamaus:2009op}.}
\label{fig:eigensys}
\end{figure}
The stochasticity matrix can be diagonalized as
\be
\sum_j \sigma_{ij} V_j^{(l)}=\lambda^{(l)} V_i^{(l)}
\ee
where $V_i^{(l)}$ are the eigenvectors and $\lambda^{(l)}$ the corresponding eigenvalues. The eigenvector corresponding to a low eigenvalue can be used as a weighting function in order to construct a halo sample that has the lowest possible stochasticity contamination \cite{Hamaus:2009op}. Furthermore, the eigenvalues allow for a clean separation of the exclusion and clustering contributions to the total noise correction.
We show the measurement and comparison to our model in Fig.~\ref{fig:eigensys}. The data show pronounced low and high eigenvalues with most of the eigenvalues identical to the fiducial shot noise. The high eigenvalue is probably related to the non-linear clustering and the low eigenvalue to exclusion. The model also predicts eight of the ten eigenvalues to agree with the fiducial shot noise as well as one high and one low eigenvalue. The exact agreement is not perfect, which is probably due to an imperfect representation of the off-diagonal stochasticity terms. The main problem with the off-diagonals is to estimate the exclusion radii. The right panel of Fig.~\ref{fig:eigensys} shows the eigenvectors corresponding to the eigenvalues. The eigenvector corresponding to the low eigenvalue is clearly connected to the mass, i.e., the exclusion volume, whereas the eigenvector corresponding to the high eigenvalue is clearly connected to the second order bias.
\par
So far we have concentrated on the stochasticity correction for narrow mass bins and quantified them in terms of the corresponding stochasticity matrix. If all the corrections were linear in the parameters of the model, all we needed to do is to calculate the corresponding mean parameters of the sample and use them to calculate the stochasticity correction for the combined sample. However, the corrections are in general a non-linear function of the  parameters. The wider the mass bins the less exact is a bulk description by a set of mean parameters. It should be more exact to consider subbins and combine them.
Thus we need to calculate the stochasticity correction for narrow mass bins $M \in [\underline{M}_i,\overline{M}_i]\ i=1,\ldots,h$. Then, when considering samples that span a wide range of halo masses or realistic galaxy samples, we need to weight the prediction for the stochasticity correction accordingly. 
The weighted density field is then 
\be
\tilde \delta=\frac{\sum_i w_i \delta_i}{\sum_i w_i}
\ee
and the corresponding noise level of the combined sample is given by
\be
\tilde \sigma=\frac{\sum_{ij}w_i \sigma_{ij}w_j}{\left(\sum_i w_i\right)^2}.
\ee
For halo samples the weighting is given by the mean number density in a mass bin
\be
w_i=\int_{\underline{M}}^{\overline{M}} \derd M n(M).
\ee
A similar weighting scheme can be derived for galaxies for the two-halo term in the context of the halo model.
\subsection{A realistic Galaxy Sample}\label{sec:realgal}
\begin{figure}[t]
\centering
\includegraphics[width=0.49\textwidth]{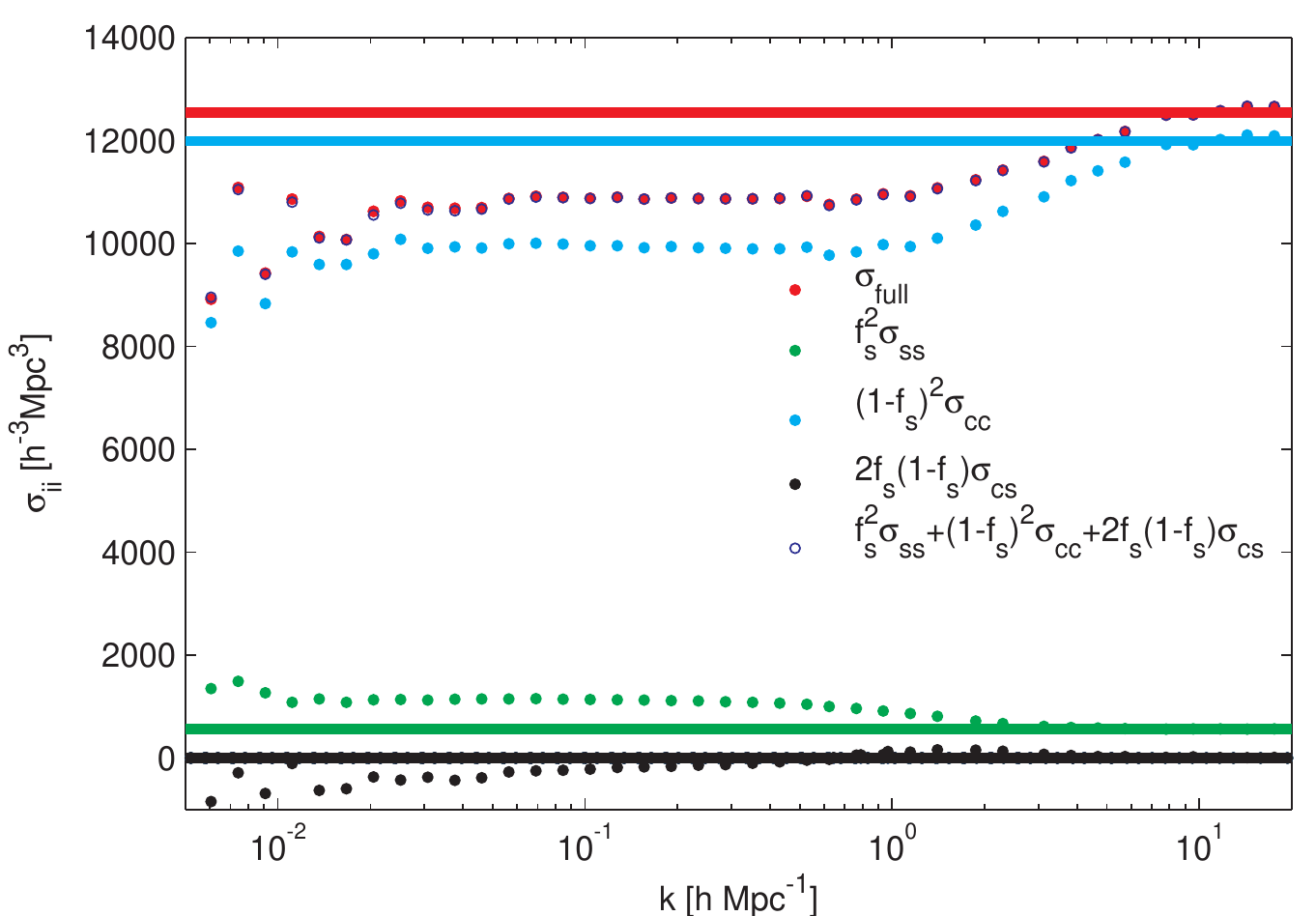}
\includegraphics[width=0.49\textwidth]{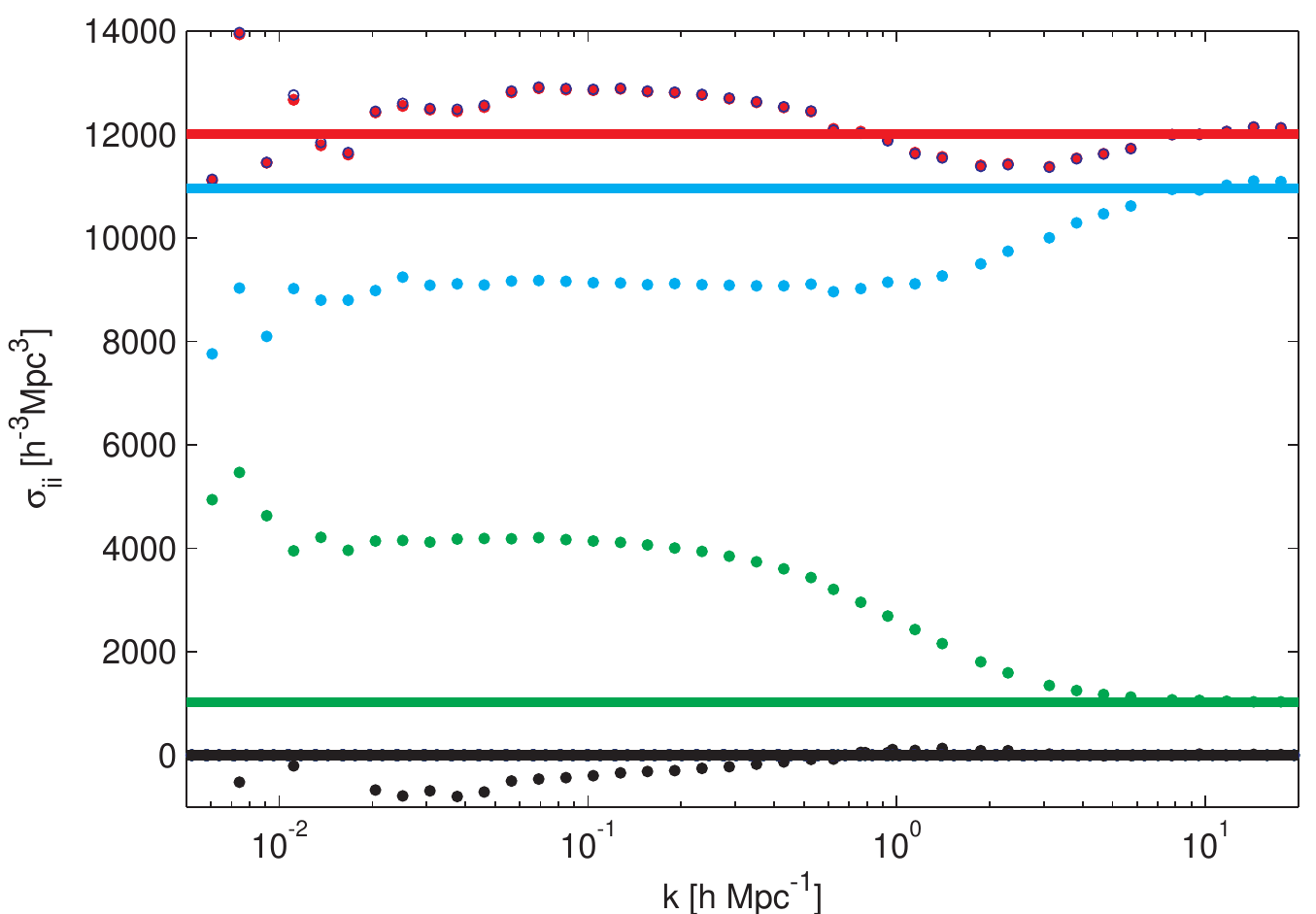}
\caption{Stochasticity of an HOD implementation of a Luminous Red Galaxy sample. We split the sample into the central-central (cyan), satellite-satellite (green) and central-satellite (black) contributions. The constituent stochasticity levels are weighted according to their contribution to the full galaxy power spectrum (see Eq.~\ref{eq:weightedsumgg}). The horizontal thick lines show the correspondingly weighted fiducial stochasticity.
\emph{Left panel: }Halo occupation distribution model with a satellite fraction of $f_\text{s}=4.9\%$ (\cite{Baldauf:2010al,Reyes:2010co}).
\emph{Right panel: }Same as left panel, but for a satellite fraction of $f_\text{s}=8.5\%$.}
\label{fig:galaxies}
\end{figure}
Let us now see how the stochasticity matrix behaves for a realistic galaxy sample. 
In Halo Occupation Distribution (HOD) models \cite{Berlind:2002th,Zheng:2005th} the occupation number $N_\text{g}(M)$ is usually split into a central and a satellite component $N_\text{g}=N_\text{c}+N_\text{s}$. In Fig.~\ref{fig:galaxies} we show the stochasticity of the Luminous Red Galaxy (LRG) sample described in \cite{Baldauf:2010al,Reyes:2010co}.
The total number density of the LRGs is $\bar n_\text{g}=7.97\tim{-5}\hMpcc$ corresponding to a fiducial shot noise of $1/\bar n_\text{g}\approx 1.25\tim{4} \hMpcc$.  The effective stochasticity level for the full sample is $\text{SN}_\text{eff}=1.09\tim{4}\hMpcc$, corresponding to a correction of $\Delta P_\text{gg}=-1.8\tim{3} \hMpcc$. The satellite fraction of the galaxy sample is $4.9 \%$.\\
Let us try to understand the total correction based on the constituent central, satellite and central-satellite cross-power spectra. The sum of these three components weighted according to Eq.~\eqref{eq:weightedsumgg} agrees with the measured stochasticity of the full sample.  
The central-central power spectrum dominates the negative stochasticity correction on large scales with a weighted correction of $(1-f_\text{s})^2\Delta P_\text{cc}=-2100 \hMpcc$
The satellite-satellite power spectrum has a positive one halo contribution on large scales that contributes a weighted correction of $f_\text{s}^2\Delta P_\text{ss}=+570 \hMpcc$
The central-satellite cross power spectrum changes sign but contributes about $\Delta P_\text{cs}=-370 \hMpcc$ at $k=0.03 \ihMpc$. 
The amplitude of these corrections could in principle be understood based on a accurate model for the stochasticity correction of the host haloes and the halo model. In this context the corrections are given as \cite{Seljak:2000an,Cooray:2002ha}
\begin{align}
P_\text{cc}^\text{(1h)}(k)=&\frac{1}{\bar n_\text{c}}\\
P_\text{ss}^\text{(1h)}(k)=&\frac{1}{\bar n_\text{s}}+\frac{1}{\bar n_\text{s}^2}\int \derd M n(M) N_\text{s,h}(M)\left[N_\text{s,h}(M)-1\right]u^2(k|M) \Theta(N_\text{s,h}-1)\\
P_\text{cs}^\text{(1h)}(k)=&\frac{1}{\bar n_\text{c}\bar n_\text{s}}\int \derd M n(M) N_\text{c,h}(M)N_\text{s,h}(M) u(k|M) \Theta(N_\text{s,h}-1)
\end{align}
\begin{align}
P_\text{cc}^\text{(2h)}(k)=&\frac{1}{\bar n_\text{c}^2} \int \derd M n(M)N_\text{c,h}(M) \int \derd M' n(M') N_\text{c,h}(M') P_\text{hh}(k|M,M')\\
P_\text{ss}^\text{(2h)}(k)=&\frac{1}{\bar n_\text{s}^2} \int \derd M n(M)N_\text{s,h}(M) u(k|M)\int \derd M' n(M') N_\text{s,h}(M') u(k|M')P_\text{hh}(k|M,M')\\
P_\text{cs}^\text{(2h)}(k)=&\frac{1}{\bar n_\text{s}\bar n_\text{c}}\int \derd M n(M)N_\text{c,h}(M) \int \derd M' n(M') N_\text{s,h}(M') u(k|M')P_\text{hh}(k|M,M')
\end{align}
On large scales we have $u(k|M)\to 1$. Furthermore the halo-halo power spectra can be again split into a linear bias part $b(M)b(M')P(k)$ and a correction term accounting for the discreteness of the host haloes.
For the central galaxy sample our model yields a correction of $\Delta P_\text{cc}\approx -1000 \hMpcc$.
More accurate predictions would require a better model of the stochasticity corrections, which in turn requires a better model of exclusion and non-linear biasing. \\
We also consider a slightly modified galaxy sample with a larger satellite fraction $f_\text{s}=8.47\%$. For this purpose we create a copy of each satellite galaxy at twice its separation from the host halo centre. The resulting stochasticity properties are shown in the right panel of 
Fig.~\ref{fig:galaxies}. In contrast to the previous case the actual stochasticity now exceeds the fiducial shot noise due to the strong positive contribution of the satellite-satellite one halo term.
\section{Conclusions}\label{sec:concl}
In this paper we discuss effects of the discreteness and non-linear clustering of haloes on their stochasticity in the 
power spectrum. The standard model for stochasticity is the Poisson shot noise model with stochasticity given as the inverse of the 
number density of galaxies, $1/\bar{n}$.
Motivated by the results in \cite{Hamaus:2009op}, we study the distribution of haloes in Lagrangian space and estimate the effect of exclusion
and non-linear clustering of protohaloes on the stochasticity. These induce corrections relative to $1/\bar{n}$ in the low-$k$ limit. 
Exclusion lowers and non-linear clustering enhances the large scale stochasticity. The total value of the large scale stochasticity depends on which of the two effects is stronger but the amplitude of the correction does not directly depend on the abundance of the sample.
These stochasticity corrections must decay to zero for high-$k$, implying they are scale dependent in the intermediate regime. The transition scale is related to either 
the exclusion scale of the halo sample under consideration or to the non-linear clustering scale. At the final time (Eulerian space) these transition scales shrink
due to the non-linear collapse, but the low-$k$ amplitude of the stochasticity agrees with Lagrangian space, as expected from mass and momentum conservation.

While the presented model can explain the observed trend of modified stochasticity at a qualitative level, the quantitative agreement is not perfect. This is related to our imperfect modelling of the Lagrangian halo correlation function with a local bias ansatz. A more realistic modelling might be possible in the full framework of peak biasing in three dimensions, as the one dimensional results in \S \ref{sec:peaks} indicate.

 We also discuss the effects of satellite galaxies,
when a galaxy sample with a non-vanishing satellite fraction is considered. In this case the stochasticity can dramatically deviate from the auxiliary $1/\bar{n}$ value on large scales. In this case one has to identify the number density of host haloes to infer the stochasticity on large scales and account for the fact that on scales below the typical scale of the satellite profile there is a transition to the fiducial Poisson shot noise of the galaxy sample.

Finally, we consider the stochasticity matrix of haloes of different mass. We show that diagoalization of this matrix 
gives rise to one eigenvaue with a low amplitude, with the eigenvector that 
aproximately scales with the halo mass.
This provides an explanation to the stochasticity suppression with mass weighting explored in \cite{Seljak:2009ho,Hamaus:2009op}. 
It would be interesting to explore, how the stochasticity corrections imprint themselves in the halo bispectrum, which is a promising probe of inflationary physics \cite{Baldauf2011a} and whose measurement becomes realistic in present and upcoming surveys \cite{Marin:2013th}.

\begin{acknowledgements}
The authors would like to thank Niayesh Afshordi, Kwan Chuen Chan, Donghui Jeong, Patrick McDonald, Teppei Okumura, Fabian Schmidt, Ravi Sheth, Zvonimir Vlah and Jaiyul Yoo for useful discussions.
The simulations were carried out on the ZBOX3 supercomputer at the Institute for Theoretical Physics of the University of Zurich.
This work is supported in part by NASA ATP Grant number NNX12AG71G, Swiss National Foundation (SNF) under contract
200021-116696/1, WCU grant R32-10130, National Science Foundation Grant No. 1066293. T.B.  acknowledges the hospitality of the Aspen Center for Physics.
\end{acknowledgements}

\appendix
\section{Perturbative Peak Effects in the Initial Correlation Function}\label{app:peakeffects}
\begin{figure}
\includegraphics[width=0.49\textwidth]{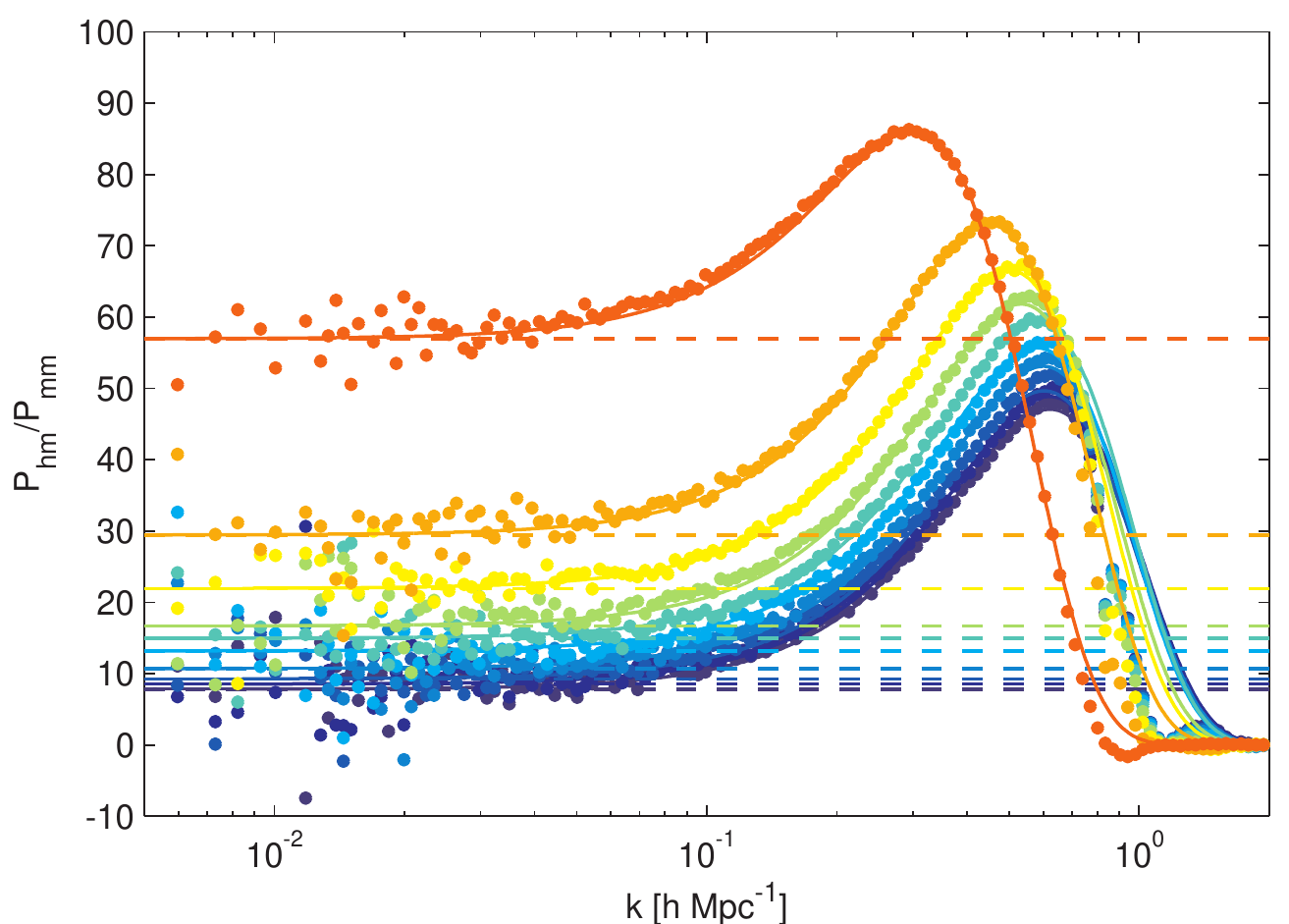}
\includegraphics[width=0.49\textwidth]{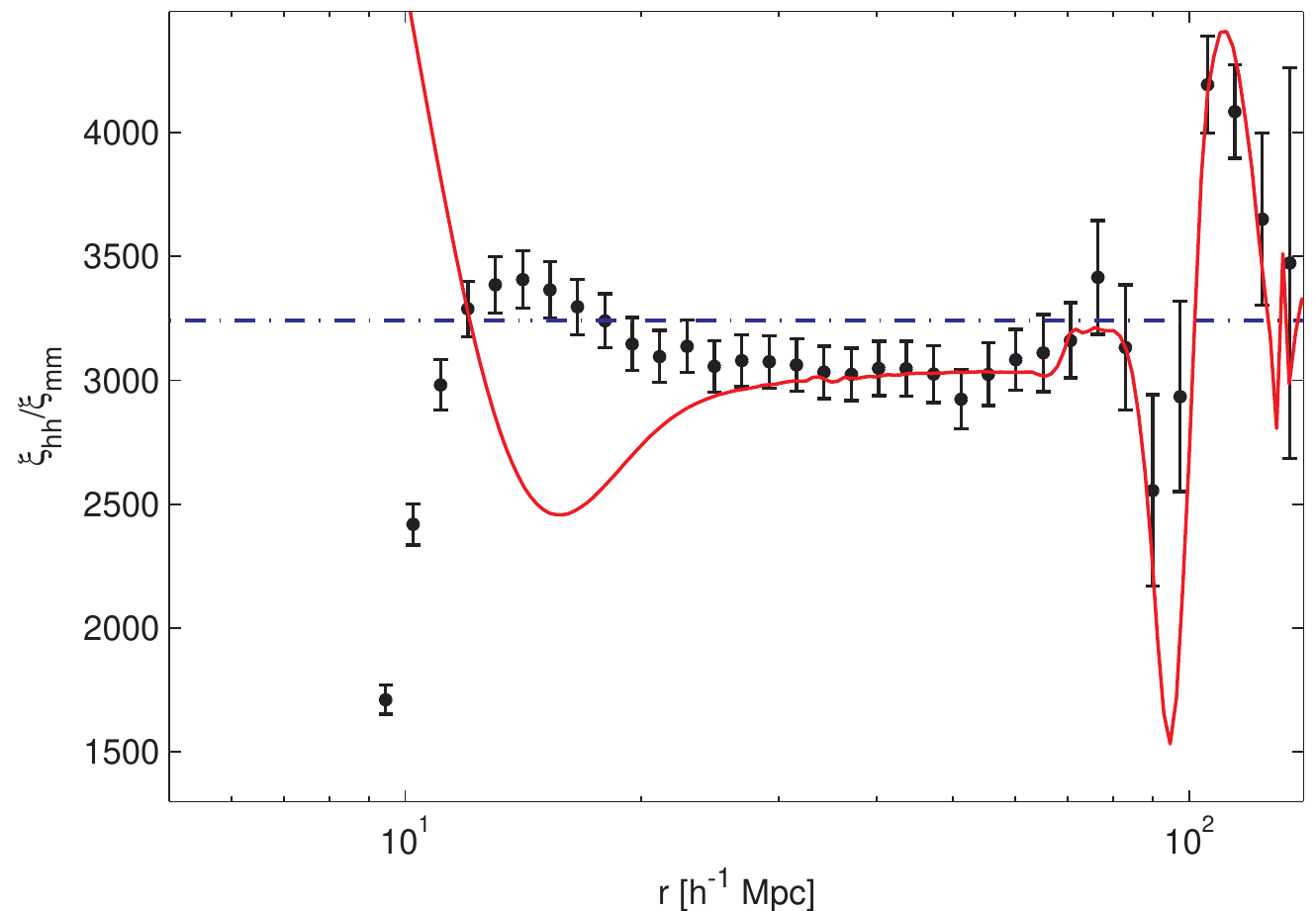}
\caption{\emph{Left panel: } Protohalo-matter cross power spectrum. Horizontal dashed lines show the linear bias and solid lines a fit using the functional form suggested by the peak model. \emph{Right panel: }Initial condition halo-halo correlation function divided by linear initial condition correlation function for mass bin X with linear scale independent bias (horizontal dash-dotted) and linear peak bias (red solid). Around the BAO scale and down to $r=30 \hMpc$ the peak model describes the ratio of halo and matter correlation functions much better, but then it fails miserably.}
\label{fig:scaledepbiasinit}
\end{figure}

As pointed out by \cite{Elia:2011th}, there is strong numerical evidence for the presence of $k^2$-terms in the linear bias of protohalo. Such terms in fact appear in all the Lagrangian bias factors predicted by the peak model, as can be seen from a large scale expansion of the peak correlation function \cite{Desjacques:2008ba,Desjacques:2010mo}. In the particular case of the peak-matter cross-correlation, the linear bias expansion is exact on all scales \cite{Desjacques:2010re} and agrees with the average density profile derived in \cite{Bardeen:1985tr}.
Under the assumption that haloes are represented by peaks, we thus have for the halo-matter power spectrum in the initial conditions
\be
P_\text{hm}(k)=b_1\left(1+\kappa_\delta k^2 \right)W_{\text{G},R_\text{pk}}(k)P(k),
\ee
where $W_{\text{G},R_\text{pk}}$ is the Gaussian filter of scale $R_\text{pk}$. Note that the peak smoothing scale is not necessarily related to the exclusion scale.
Similarly one can expand the peak-peak correlation function on large scales and obtain a scale dependent linear bias relation that would relate to the halo-halo power spectrum of the form
\be
P_\text{hh}(k)\approx b_1^2 \left(1+\kappa_\delta k^2 \right)^2W^2_{\text{G},R_\text{pk}}(k)P(k).
\ee
One has to be careful when considering the correspondence between large scales in the correlation function and small wavenumbers in the power spectrum. As we stressed above, the amplitude of the low-$k$ power spectrum is tightly coupled to the small scale correlation function. Thus, there is no reason to believe that the large scale expansion of the correlation function will yield a correct description of the low-$k$ power spectrum. As we will discuss in more detail below, a correct description of the low-$k$ power spectrum in the peak model requires higher order bias expansions or a non-perturbative evaluation of the full peak correlation function.\\
Given the functional form of the halo-matter power spectrum in the initial conditions we can fit for the linear bias $b_1$, the relative peak bias $\kappa_\delta$ and the smoothing scale $R_\text{pk}$. We show the corresponding scale dependent bias in the left panel of Fig.~\ref{fig:scaledepbiasinit}. This plot also shows the linear local (i.e. scale independent) bias parameter used in the main text as the horizontal dashed line.\\
We can use the inferred bias parameters and peak smoothing scale to calculate the corresponding imprint on the halo-halo correlation function. As we show in the right panel of Fig.~\ref{fig:scaledepbiasinit}, the correlation function deviates significantly from the na{\"{\i}}ve linear scale independent bias prediction shown as the horizontal dash-dotted line. On scales exceeding $r\approx 25\hMpc$, the measured correlation function is in much better correspondence with the linear peak bias shown as the solid red line. Below $25 \hMpc$ the peak correlation fails to predict the scale dependence and does actually worse than the linear scale independent bias. This had to be expected, since the linear peak bias is only accurate on large scales and on smaller scales higher order bias corrections need to be taken into consideration. In fact, comparison between the linear peak bias and the full numerical evaluation of the one dimensional peak model, suggests that $20 - 30 \hMpc$ is a typical breakdown scale for the linear peak bias. We will explore the convergence properties of the perturbative peak model in more detail in a forthcoming paper.\\
While accurately predicting the functional form of the scale dependent bias, the mass dependence of the coefficients deviates from the predictions of the peak model. The main reason for discussing the peak corrections here, is that the presence of the peak corrections invalidates the simple $b_2$ fitting procedure following Eq.~\eqref{eq:xibiassecond}. In this simple approach we consider the linear bias from the low-$k$ limit of the cross power spectrum and consider the positive correction $b_2^2 \xi^2$ on top of it. The right panel of Fig.~\ref{fig:scaledepbiasinit} rather suggests that the quadratic bias corrections need to be considered on top of the linear peak bias. Furthermore the peak model predicts several second order bias contributions in contrast to the one bias contribution arising from the local model. Fitting the second order peak bias parameters would require higher order spectra, such as the protohalo bispectrum and even the second order biasing might not be sufficient to explain the upturn in the halo-halo correlation function. We thus restrict ourselves to the simple local model and stress that the quadratic bias parameters are a phenomenological fit rather than a true quadratic bias and we thus don't expect them to be in accordance with the peak-background split prediction.
\section{The One Dimensional Peak Model}\label{app:1dpeaks}
Here we review the major  steps in the derivation of the one-dimensional peak model following \cite{Lumsden:1989th}.
As mentioned before, the peak model associates maxima of the density fields with the formation sites of dark matter haloes. Thus, we are interested in the clustering statistics of these points. For simplicity we will only consider a one dimensional field, which would be for a example a skewer through the full three dimensional cosmological density field.
The number density of peaks can be written as a sum over delta functions at the peak positions $x_\text{pk}$
\be
n_\text{pk}(x)=\sum_\text{pk}\delta^\text{(D)}(x- x_\text{pk})
\ee
and $\delta_\text{pk}(x)=n_\text{pk}(x)/\bar n_\text{pk}-1$.
Using that the first derivative of the field vanishes at the peak position, we can expand the density field around the peak position
\be
\delta(x)\approx\delta(x_\text{pk})+\frac12 \delta''(x_\text{pk})(x-x_{\text{pk}})^2.
\ee
Taking the derivative, we obtain
\be
\delta'( x)\approx\delta''( x_\text{pk})(x-x_{\text{pk}}).
\ee
Using the transformation properties of the Dirac delta we have
\be
\delta^\text{(D)}( x- x_\text{pk})=\delta''(x_\text{pk}) \delta^\text{(D)}( \delta').
\ee
This expression is known as the Kac-Rice formula \cite{Desjacques:2008ba,Adler:2007}.
The mean number density is readily obtained as an integral over the one point probability density function of the field amplitude, slope and curvature
\be
\bar n_\text{pk}=\la n_\text{pk}( x)\ra=\int \derd \vec y\  \mathbb P_\text{1pt}(\vec y) \delta'' \ddir(\delta')
\label{eq:peaknum}
\ee
where $\vec y=(\delta, \delta',\delta'')$. The two point correlation function is then given by
\begin{align}
\la \delta_\text{pk}(x_1)\delta_\text{pk}(x_2)\ra=&\int \derd \vec Y\ \mathbb P_\text{2pt}(| x_1-x_2|;\vec Y) \delta_\text{pk}(x_1)\delta_\text{pk}(x_2)\\
=&\frac{1}{{\bar n_\text{pk}}^2}\int \derd \vec Y\ \mathbb P_\text{2pt}(|x_1- x_2|;\vec Y) \delta''_1 \delta''_2 \ddir(\delta'_1) \ddir(\delta'_2)-1,
\label{eq:peakcorrel}
\end{align}
where $\vec Y=(\delta_1,\delta'_1,\delta''_1,\delta_2,\delta'_2,\delta''_2)$. The one and two point PDFs are given by
\begin{align}
\mathbb P_\text{1pt}(\vec y)=\frac{1}{\sqrt{(2\pi)^3\det m}}\eh{-\frac12 y m^{-1} y^T}, && \mathbb P_\text{2pt}(\vec Y)=\frac{1}{\sqrt{(2\pi)^6\det M}}\eh{-\frac12 Y M^{-1} Y^T}.
\end{align}
The symmetric $6\times6$ covariance matrix of the field amplitude and derivatives $M_{ij}=\la Y_i Y_j \ra$ can then be written as
\be
M
=\begin{pmatrix}
m & B(r)\\
B^\text{T}(r) & m
\end{pmatrix},
\label{eq:blockmatrix}
\ee
where the constituent block matrices are given by
\begin{align}
m=\begin{pmatrix}
\sigma_0^2& 0 & -\sigma_1^2\\
0 & \sigma_1^2 & 0\\
-\sigma_1^2 & 0 & \sigma_2^2
\end{pmatrix},
&&
B(r)=\begin{pmatrix}
\xi_0(r)& -\xi_{1/2}(r) & -\xi_1(r)\\
\xi_{1/2}(r) & \xi_1(r) & -\xi_{3/2}(r)\\
-\xi_1(r) & \xi_{3/2}(r) & \xi_2(r)
\end{pmatrix}.
\end{align}
The only remaining ingredient for the evaluation of Eq.~\eqref{eq:peakcorrel} are the correlators of field amplitudes and derivatives, which we obtain by smoothing the three dimensional density field with a Gaussian filter and considering its values and derivatives along one coordinate axis, which we choose to be the $z$-direction without loss of generality
\be
\xi_{(n+m)/2}(r)=\int \frac{\derd^3 k}{(2\pi)^3} (-1)^n (\ii \mu k)^{n + m} \eh{\ii \mu k r} P(k)W_{\text{G},R_\text{pk}}(k),
\ee
where $\mu=\hat{\vec k}\cdot \hat{\vec z}$. The moments of field amplitudes and derivatives are then given as $\sigma_{(n+m)/2}^2=\xi_{(n+m)/2}(0)$.
\section{The Exclusion Kernel}\label{app:exclkern}
Let us assume, that there is some scatter around the mean exclusion radius $R$ of some sample due to triaxiality and a finite width of the mass bin. Let us furthermore assume that this scatter is lognormally distributed such that the PDF is given by
\be
f(r)=\frac{1}{x\sigma \sqrt{2\pi}}\eh{-\frac{\ln^2(r/R)}{2\sigma^2}}.
\ee
Let us now calculate the probability of finding a pair with an actual separation that is smaller than the scale under consideration
\be
F(r)=\int_0^r \derd r' f(r')=\frac{1}{2}\left(1+\text{erf}\left[\frac{\ln(r/R)}{\sqrt{2}\sigma}\right]\right).
\ee
To find a pair at separation $r$ its exclusion scale needs to be smaller than the actual separation and the total probability is given as a product of the fiducial probability of finding a pair in a non-excluded sample and the probability that the actual exclusion scale is smaller than $r$
\be
\delta\tilde{\mathbb P}(r)=F(r)\delta\mathbb{P}(r)=\bar n\left[1+\xi^\text{(d)}(r)\right]\delta V=F(r)\bar n \left[1+\xi^\text{(c)}(r)\right]\delta V.
\ee
We can now infer the correlation of the discrete tracers
\be
\xi^\text{(d)}(r)=F(r)\left[1+\xi^\text{(c)}(r)\right]-1
\ee
In the main text we use the logarithm to base 10 for convenience rather than the natural logarithm. The corresponding scatters are related by a simple rescaling by a factor of $\log_{10} e \approx0.43$.

\begin{figure}[p]
\includegraphics[height=0.99\textheight]{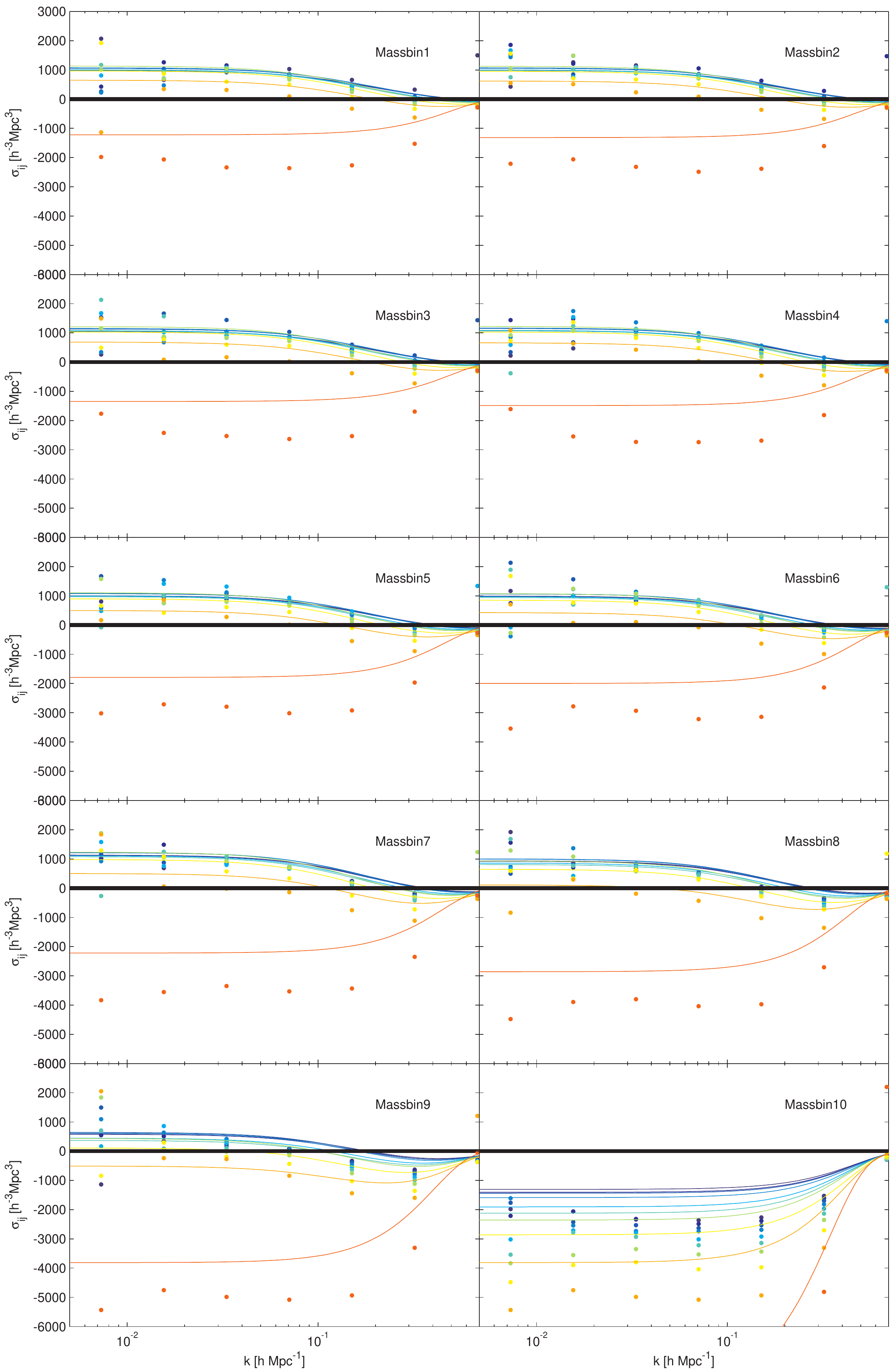}
\caption{Cross-terms of the stochasticity matrix of the $z=0$ haloes traced back to the initial conditions. The text in the panel refers to the mass bin with which we cross correlate all the other bins. Note that for reference we also include the diagonals of the matrix with the fiducial shot noise subtracted out. Mass increases from top to bottom, i.e., blue to orange.}
\label{fig:crossterms}
\end{figure}

\bibliography{shotnoise.bib}

\end{document}